\documentclass[aps,pre,preprint,superscriptaddress,showpacs]{revtex4-1}

\usepackage{amsmath, amsfonts, amsthm, amssymb, mathrsfs, enumerate, graphicx, cases, hyperref}
\usepackage[T1]{fontenc}

\begin{document}


\title{Mass-loss of an isolated gravitating system due to energy carried away by gravitational waves with a cosmological constant}


\author{Vee-Liem Saw}
\email[]{VeeLiem@maths.otago.ac.nz}
\affiliation{Department of Mathematics and Statistics, University of Otago, Dunedin 9016, New Zealand}


\date{\today}

\begin{abstract}
We derive the asymptotic solutions for vacuum spacetimes with non-zero cosmological constant $\Lambda$, using the Newman-Penrose formalism. Our approach is based exclusively on the physical spacetime, i.e. we do not explicitly deal with conformal rescaling nor the conformal spacetime. By investigating the Schwarzschild-de Sitter spacetime in spherical coordinates, we subsequently stipulate the fall-offs of the null tetrad and spin coefficients for asymptotically de Sitter spacetimes such that the terms which would give rise to the Bondi mass-loss due to energy carried by gravitational radiation (i.e. involving $\sigma^o$) must be non-zero. After solving the vacuum Newman-Penrose equations asymptotically, we propose a generalisation to the Bondi mass involving $\Lambda$ and obtain a positive-definite mass-loss formula by integrating the Bianchi identity involving $D'\Psi_2$ over a compact 2-surface on $\mathcal{I}$. Whilst our original intention was to study asymptotically de Sitter spacetimes, the use of spherical coordinates implies that this readily applies for $\Lambda<0$, and yields exactly the known asymptotically flat spacetimes when $\Lambda=0$. In other words, our asymptotic vacuum solutions with $\Lambda\neq0$ reduce smoothly to those where $\Lambda=0$, in spite of the distinct characters of $\mathcal{I}$ being spacelike, timelike and null for de Sitter, anti-de Sitter and Minkowski, respectively. Unlike for $\Lambda=0$ where no incoming radiation corresponds to setting $\Psi^o_0=0$ on some initial null hypersurface, for $\Lambda\neq0$, no incoming radiation requires $\Psi^o_0=0$ everywhere.
\end{abstract}


\maketitle


\section{Introduction}

Back in the early days of general relativity, the question of whether gravitational wave is a genuine physical phenomenon or merely a coordinate/gauge artefact was a major research problem. Albert Einstein himself solved his field equations to first order and found a wave-like solution with the quadrupole formula, just a year after he formulated them in 1915 \cite{einquad}. Two decades later with Nathan Rosen, they solved the full field equations without approximations and initially concluded that gravitational radiation did not exist as their solution involved singularities. Upon the realisation that this was merely a coordinate singularity, Einstein later retracted, and today that solution is known as the Einstein-Rosen cylindrical gravitational waves \cite{EinPhyRev,EinRosss}. Whilst many physicists circa 1960 were by then convinced about the physical nature of gravitational waves \footnote{According to the sticky bead argument \cite{bead}, a propagating gravitational wave would cause a bead on a stick held transverse to the wave propagation to oscillate. Friction between the bead and the stick would cause them to heat up. This implies that gravitational wave must carry energy away from its source, which is then imparted to the sticky bead.}, this fundamental question still needed a rigorous theoretical framework. The observational front provided no significant support either, as gravitational radiation was too feeble to be detected. It was only this year, that LIGO announced its direct detections \cite{LIGO,LIGO2}, after the indirect observation of the decrease in orbital period of the Hulse-Taylor binary system in 1974 due to energy loss in emitting gravitational radiation \cite{HT1,HT2,HT3,HT4,HT5}.

A noteworthy breakthrough was the calculations by Bondi, who worked out in the case of axisymmetry, that when an isolated gravitating system emits gravitational radiation, its total mass-energy would decrease. Hence, that was a first useful conclusive definition of mass-energy, excluding the energy carried by gravitational waves (which we now refer to as the Bondi mass) \cite{Bondi60,Bondi62} \footnote{There is a different definition of mass-energy, called the ADM mass which does include the energy of gravitational radiation. So this remains constant even though the system emits gravitational radiation \cite{adm,Poisson}.}.

Shortly after, a complete treatment for the full Einstein theory (i.e. for general spacetimes, without assuming axisymmetry, etc.) was carried out by Newman and Unti \cite{newunti62} by solving the Newman-Penrose equations \cite{newpen62}. These equations are equivalent to the Einstein field equations, essentially rewriting and re-expressing the Christoffel symbols (connection coefficients) in terms of some complex quantities called \emph{spin coefficients}. Most of these equations are linear. They were thus able to solve this asymptotically where $r\rightarrow\infty$. One of the Bianchi identities involving the derivative along $\mathcal{I}$ of the dyad component of the Weyl spinor, $D'\Psi_2$, integrated over a compact 2-surface at infinity, would correspond exactly to Bondi's result. With $M_B$ denoting the Bondi mass-energy of an isolated gravitating system, that formula is:
\begin{eqnarray}\label{one}
\frac{dM_B}{du}=-\frac{1}{A}\oint{|\dot{\sigma}^o|^2d^2S},
\end{eqnarray}
where here, $A = 4\pi$ (the area of the unit sphere).  A non-zero $\sigma^o$ is interpreted as the presence of gravitational waves being emitted by the system. Dot is differentiation with respect to $u$, the retarded null coordinate. Ergo, this equation says that the Bondi mass of a system drops whenever gravitational waves are emitted. The result is hailed as a milestone, because it signifies that gravitational waves constitute a real physical phenomenon, carrying energy away from the system and reducing the system's total mass-energy. This is all good, where the cosmological constant is set to zero.

Latterly, we know that the rate of expansion of our universe is accelerating \cite{cosmo1,cosmo2}. This accelerating rate of expansion may be explained very well by sticking in a positive cosmological constant $\Lambda$ in the Einstein field equations. Since this is a recent discovery, the problem of extending the Bondi mass, and obtaining the Bondi mass-loss formula (due to energy carried away by emitting gravitational waves) with $\Lambda>0$ is of great current interest. Within the past couple of years (note that Refs. \cite{Szabados,ash1,ash2,ash3,ash4,chi1,chi2,Chrusciel,gracos1,gracos2} are from late 2014 onwards), there has been a raft of ongoing research with different approaches. After reviewing the conformal properties of de Sitter-like spacetimes, Szabados and Tod worked towards a positive Bondi-type mass using twistor methods and the Nester-Witten two-form \cite{Szabados}. A series of papers by Ashtekar et al. \cite{ash1,ash2,ash3,ash4} began by reporting that it is necessary for the conformal boundary $\mathcal{I}$ of de Sitter-like spacetimes to be non-conformally flat if gravitational radiation emitted by isolated systems would carry energy away from the source \cite{ash1}, and proceeded to study the linearised theory \cite{ash2} as well as derive the corresponding quadrupole formula \cite{ash3} with $\Lambda>0$. Interestingly, $\Lambda>0$ allows for arbitrarily negative energy carried by the gravitational waves, though they argued that physically reasonable sources would radiate waves of positive energy. Apart from that, He et al. followed Bondi's original study on axisymmetric isolated systems \cite{Bondi62} to calculate the dyad component of the Weyl spinor $\Psi_4$ which represents outgoing radiation \cite{chi1}, and subsequently extended it to couple with Maxwell's equations \cite{chi2}. They succeeded by generalising Bondi's boundary condition to the leading order of the metric ansatz for non-zero $\Lambda$, which incidentally corresponds to a non-conformally flat $\mathcal{I}$. Other approaches to this problem include those by Chru\'{s}ciel and Ifsits involving the notion of renormalised volume \cite{Chrusciel}, as well as linearisations by Refs. \cite{gracos1,gracos2}.

In this paper, we adopt the approach by Newman-Unti \cite{newunti62} and solve the Newman-Penrose equations \cite{newpen62} with non-zero $\Lambda$ \cite{don}, asymptotically. This involves solving 38 equations comprising the metric equations, spin coefficient equations and the Bianchi identities. In contrast to Newman and Penrose who made minimal assumptions on the differentiability criteria of the relevant quantities when they did this for $\Lambda=0$ \cite{newpen62}, we shall assume that the null tetrad, spin coefficients and the dyad components of the Weyl spinor all have series expansions away from $\mathcal{I}$ of sufficiently many orders. Moreover, as we shall be working with spherical coordinates, these calculations would also apply to anti-de Sitter-like spacetimes for $\Lambda<0$, and reduce to asymptotically flat spacetimes for $\Lambda=0$.

We proceed in Section 2 to discuss the Schwarzschild-de Sitter spacetime in spherical coordinates, working out the null tetrad, spin coefficients, and the Weyl spinor. This would serve as a good exercise as well as illustrating what the general leading terms for the fall-offs would probably look like. Then in the following section, we stipulate the fall-offs for asymptotically de Sitter spacetimes. Perhaps unexpectedly (unless one studies the conformally rescaled spacetime, for instance --- see Ref. \cite{Szabados}), two spin coefficients, viz. $\sigma'$ and $\kappa'$ mandatorily require non-zero terms of order $O(1)$, otherwise the usual shear term $\sigma^o$ which represents the Bondi news would necessarily vanish \footnote{We actually discovered this whilst in the process of solving the 38 Newman-Penrose equations, to find that $\sigma^o=0$ without those $O(1)$ terms in $\sigma'$ or $\kappa'$. These two spin coefficients vanish for spherically symmetric spacetimes since they have non-zero spin-weights, so the Schwarzschild-de Sitter spacetime gave no prior warning --- incongruous to what we would have hoped for in trying to guess the fall-offs.}. Once these have been solved asymptotically (with details on solving these 38 equations order-by-order found in the appendix), we present the summary of our extensive calculations for the asymptotic behavior of vacuum spacetimes with non-zero $\Lambda$ in Section 4, with a proposed generalisation of the Bondi mass and the resulting mass-loss formula in Section 5 by integrating the Bianchi identity involving $D'\Psi_2$ over a compact 2-surface of constant $u$ on $\mathcal{I}$. Section 6 deals more closely with these 2-surfaces, where we also look at the specialisation with axisymmetry. One can calculate the Cotton tensor for $\mathcal{I}$ to find that it is generally non-zero unless $\sigma^o=0$, i.e. $\mathcal{I}$ is non-conformally flat when energy is carried away from the isolated source due to gravitational radiation. Finally, we discuss these new physics due to the cosmological constant in Section 7. We follow the notations and conventions of Newman and Penrose, which are in line with Refs. \cite{Pen87,Pen88}. The constants $c$ and $G$ are set to 1.

\newpage

\section{Schwarzschild-de Sitter spacetime}

The Schwarzschild-de Sitter spacetime can be expressed in spherical coordinates \cite{GP}:
\begin{eqnarray}\label{deSittermetric}
g=J(r)dt^2-\frac{1}{J(r)}dr^2-r^2d\Omega^2,
\end{eqnarray}
where $J(r)=1-\Lambda r^2/3-2M/r$. The constant $M$ is the mass, with $\Lambda$ being the cosmological constant ($\Lambda>0$, $\Lambda<0$, and $\Lambda=0$ correspond to de Sitter anti-de Sitter, and Minkowski spacetimes, respectively, when $M=0$). The unit 2-sphere is $d\Omega^2=d\theta^2+\sin^2{\theta}d\phi^2$. The coordinates take values $-\infty<t<\infty$, $0\leq\theta\leq\pi$, $0\leq\phi<2\pi$, and $0<r<\infty$. Note that for $\Lambda>0$ and $M=0$, the interval $0\leq r<\sqrt{3/\Lambda}$ implies that $r$ and $t$ are spacelike and timelike coordinates, respectively, whereas the interval $r>\sqrt{3/\Lambda}$ implies that $r$ and $t$ are timelike and spacelike coordinates, respectively. (For $M\neq0$, there would be another value of $r$, viz. the event horizon of the black hole where the $r$ and $t$ coordinates would interchange their causal characters.)

To carry out the asymptotic expansion for de Sitter-like spacetimes \`{a} la Newman-Unti \cite{newunti62}, it is desirable to express the de Sitter metric (when $M=0$, though we shall for the moment do it generally for the Schwarzschild-de Sitter metric) in the form
\begin{eqnarray}
g=g_{uu}(u,r)du^2+2dudr-r^2d\Omega^2,
\end{eqnarray}
where $u$ would be a retarded null coordinate (usually related to $t$ and $r$ via the relation of the form ``$u=t-r$''). Let $r_*=\int{dr/J(r)}$, $dr_*=dr/J(r)$, $J(r)dr_*^2=dr^2/J(r)$. Then,
\begin{eqnarray}
g=J(r)(dt^2-dr_*^2)-r^2d\Omega^2.
\end{eqnarray}
Furthermore, let $u=t-r_*$, $du=dt-dr_*$, $dt^2=dr_*^2+du^2+2dudr_*$, so
\begin{eqnarray}
g&=&J(r)du^2+2J(r)dudr_*-r^2d\Omega^2\\
&=&J(r)du^2+2dudr-r^2d\Omega^2\\
&=&-\left(\frac{\Lambda}{3}r^2-1+\frac{2M}{r}\right)du^2+2dudr-r^2d\Omega^2.
\end{eqnarray}

The inverse metric has components
\begin{eqnarray}
g^{ab}=
\begin{pmatrix}
  0 & 1 & 0 & 0 \\
  1 & \frac{\Lambda}{3}r^2-1+\frac{2M}{r} & 0 & 0 \\
  0 & 0 & -\frac{1}{r^2} & 0 \\
	0 & 0 & 0 & -\frac{1}{r^2}\csc^2{\theta}
\end{pmatrix}.
\end{eqnarray}
Let $\tilde{l}=\tilde{d}u$, so we can define a null tetrad
\begin{eqnarray}
\vec{l}&=&\vec{\partial}_r\label{tet1}\\
\vec{n}&=&\vec{\partial}_u+\left(\frac{\Lambda}{6}r^2-\frac{1}{2}+\frac{M}{r}\right)\vec{\partial}_r\\
\vec{m}&=&\frac{1}{\sqrt{2}r}\vec{\partial}_{\theta}+\frac{i}{\sqrt{2}r}\csc{\theta}\vec{\partial}_{\phi}\\
\vec{\bar{m}}&=&\frac{1}{\sqrt{2}r}\vec{\partial}_{\theta}-\frac{i}{\sqrt{2}r}\csc{\theta}\vec{\partial}_{\phi},\label{tet2}
\end{eqnarray}
with $g^{ab}=l^an^b+n^al^b-m^a\bar{m}^b-\bar{m}^am^b$. The directional derivatives are
\begin{eqnarray}
D&=&l^a\nabla_a=\frac{\partial}{\partial r}\\
D'&=&n^a\nabla_a=\frac{\partial}{\partial u}+\left(\frac{\Lambda}{6}r^2-\frac{1}{2}+\frac{M}{r}\right)\frac{\partial}{\partial r}\\
\delta&=&m^a\nabla_a=\frac{1}{\sqrt{2}r}\frac{\partial}{\partial\theta}+\frac{i}{\sqrt{2}r}\csc{\theta}\frac{\partial}{\partial\phi}\\
\delta'&=&\bar{m}^a\nabla_a=\frac{1}{\sqrt{2}r}\frac{\partial}{\partial\theta}-\frac{i}{\sqrt{2}r}\csc{\theta}\frac{\partial}{\partial\phi}.
\end{eqnarray}
In the Newman-Penrose formalism, the Christoffel symbols (connection coefficients) are re-expressed in terms of twelve complex spin coefficients, defined as \cite{Pen87,don}
\begin{eqnarray}
\kappa&=&m^aDl_a\\
\kappa'&=&\bar{m}^aD'n_a\\
\sigma&=&m^a\delta l_a\\
\sigma'&=&\bar{m}^a\delta'n_a\\
\tau&=&m^aD'l_a\\
\tau'&=&\bar{m}^aDn_a\\
\rho&=&m^a\delta'l_a\\
\rho'&=&\bar{m}^a\delta n_a\\
\gamma&=&\frac{1}{2}(n^aD'l_a-\bar{m}^aD'm_a)\\
\gamma'&=&\frac{1}{2}(l^aDn_a-m^aD\bar{m}_a)\\
\alpha&=&\frac{1}{2}(n^a\delta'l_a-\bar{m}^a\delta'm_a)\\
\alpha'&=&\frac{1}{2}(l^a\delta n_a-m^a\delta\bar{m}_a).
\end{eqnarray}
The commutator equations acting on these spin coefficients are \cite{Pen87,don}
\begin{eqnarray}
[D',D]&=&2\textrm{Re}(\gamma)D-2\textrm{Re}(\gamma')D'+(\tau'-\bar{\tau})\delta+(\bar{\tau}'-\tau)\delta'\\
{[}\delta,D]&=&(\bar{\tau}'+\bar{\alpha}-\alpha')D+\kappa D'+(2i\textrm{Im}(\gamma')-\bar{\rho})\delta-\sigma\delta'\\
{[}\delta,D']&=&\bar{\kappa}'D+(\tau+\alpha'-\bar{\alpha})D'-(2i\textrm{Im}(\gamma)+\rho')\delta-\bar{\sigma}'\delta'\\
{[}\delta',\delta]&=&2i\textrm{Im}(\rho')D-2i\textrm{Im}(\rho)D'+(\alpha+\bar{\alpha}')\delta-(\bar{\alpha}+\alpha')\delta'.
\end{eqnarray}
These spin coefficients can be solved by applying the commutators to the four coordinate functions $u,r,\theta, \phi$. The directional derivatives acting on the coordinate functions are
\begin{eqnarray}
\begin{matrix}
  Du=0, & D'u=1, & \delta u=0, & \delta'u=0, \\
  Dr=1, & D'r=\frac{\Lambda}{6}r^2-\frac{1}{2}+\frac{M}{r}, & \delta r=0, & \delta'r=0, \\
  D\theta=0, & D'\theta=0, & \delta\theta=\frac{1}{\sqrt{2}r}, & \delta'\theta=\frac{1}{\sqrt{2}r}, \\
	D\phi=0, & D'\phi=0, & \delta\phi=\frac{i}{\sqrt{2}r}\csc{\theta}, & \delta'\phi=-\frac{i}{\sqrt{2}r}\csc{\theta}.
\end{matrix}
\end{eqnarray}
The relevant non-zero second derivatives are
\begin{eqnarray}
DD'r&=&\frac{\Lambda}{3}r-\frac{M}{r^2}\\
D\delta\theta&=&-\frac{1}{\sqrt{2}r^2}\\
D\delta\phi&=&-\frac{i}{\sqrt{2}r^2}\csc{\theta}\\
D'\delta{\theta}&=&-\frac{\Lambda}{6\sqrt{2}}+\frac{1}{2\sqrt{2}r^2}-\frac{M}{\sqrt{2}r^3}\\
D'\delta{\phi}&=&\left(-\frac{\Lambda}{6\sqrt{2}}+\frac{1}{2\sqrt{2}r^2}-\frac{M}{\sqrt{2}r^3}\right)i\csc{\theta}\\
\delta'\delta\phi&=&-\frac{i}{2r^2}\csc{\theta}\cot{\theta}\\
\delta\delta'\phi&=&\frac{i}{2r^2}\csc{\theta}\cot{\theta}.
\end{eqnarray}
\newpage

Applying the commutator equations to the coordinates $u, r, \theta, \phi$ then gives
\begin{eqnarray}
[D',D]u &:& \textrm{Re}(\gamma')=0\\
{[}\delta,D]u &:& \kappa=0\\
{[}\delta,D']u &:& \tau=\bar{\alpha}-\alpha'\\
{[}\delta',\delta]u &:& \textrm{Im}(\rho)=0\\
{[}D',D]r &:& \textrm{Re}(\gamma)=-\frac{\Lambda}{6}r+\frac{M}{2r^2}\\
{[}\delta,D]r &:& \bar{\tau}'=\alpha'-\bar{\alpha}\\ 
{[}\delta,D']r &:& \kappa'=0\\
{[}\delta',\delta]r &:& \textrm{Im}(\rho')=0\\
{[}D',D]\theta &:& \textrm{Re}(\tau-\bar{\tau}')=0\\
{[}D',D]\phi &:& \textrm{Im}(\tau-\bar{\tau}')=0\\
{[}\delta,D]\theta &:& \rho+\sigma-2i\textrm{Im}({\gamma'})=-\frac{1}{r}\\
{[}\delta,D]\phi &:& \rho-\sigma-2i\textrm{Im}({\gamma'})=-\frac{1}{r}\\
{[}\delta,D']\theta &:& \rho'+\bar{\sigma}'+2i\textrm{Im}({\gamma})=-\frac{\Lambda}{6}r+\frac{1}{2r}-\frac{M}{r^2}\\
{[}\delta,D']\phi &:& \rho'-\bar{\sigma}'+2i\textrm{Im}({\gamma})=-\frac{\Lambda}{6}r+\frac{1}{2r}-\frac{M}{r^2}\\
{[}\delta',\delta]\theta &:& \textrm{Im}(\alpha')=\textrm{Im}(\alpha)\\
{[}\delta',\delta]\phi &:& \textrm{Re}(\alpha+\bar{\alpha}')=-\frac{1}{\sqrt{2}r}\cot{\theta}.
\end{eqnarray}
Well, $[\delta,D]\theta$ and $[\delta,D]\phi$ imply that $\rho=-1/r$, $\sigma=0$, $\textrm{Im}(\gamma')=0$, with $[\delta,D']\theta$ and $[\delta,D']\phi$ giving that $\rho'=-\Lambda r/6+1/2r-M/r^2$, $\sigma'=0$, $\textrm{Im}(\gamma)=0$. Next, $[D',D]\theta$ and $[D',D]\phi$ imply that $\tau=\bar{\tau}'$. Together with $\tau=\bar{\alpha}-\alpha'$ and $\bar{\tau}'=\alpha'-\bar{\alpha}$, we get $\alpha'=\bar{\alpha}$ so that $\tau=0$ and $\tau'=0$. From $[\delta',\delta]\theta$, this requires that $\textrm{Im}(\alpha)=0$ and $\textrm{Im}(\alpha')=0$. Finally, $[\delta',\delta]\phi$ leads to $\alpha=\alpha'=-\cot{\theta}/2\sqrt{2}r$.

Ergo, $\kappa=0$, $\kappa'=0$, $\sigma=0$, $\sigma'=0$, $\tau=0$, $\tau'=0$, $\gamma'=0$,
\begin{eqnarray}
\gamma&=&-\frac{\Lambda}{6}r+\frac{1}{2}Mr^{-2}\\
\rho&=&-r^{-1}\\
\rho'&=&-\frac{\Lambda}{6}r+\frac{1}{2}r^{-1}-Mr^{-2}\\
\alpha&=&\alpha'=-\frac{1}{2\sqrt{2}}\cot{\theta}\ r^{-1}.
\end{eqnarray}
We remark that the mass appears in the order of $r^{-2}$ in $\gamma$ and $\rho'$.

The Einstein field equations relate part of the spacetime curvature (viz. the Einstein tensor) to the stress-energy tensor. The remaining part of the (Riemann) curvature is contained in the Weyl tensor. In the Newman-Penrose formalism, it is convenient to introduce the Weyl spinor instead, with the dyad components (also referred to as the Weyl scalars) defined as \cite{Pen87,don}
\begin{eqnarray}
\Psi_0&=&C_{abcd}l^am^bl^cm^d\\
\Psi_1&=&C_{abcd}l^am^bl^cn^d\\
\Psi_2&=&C_{abcd}l^am^b\bar{m}^cn^d\\
\Psi_3&=&C_{abcd}l^an^b\bar{m}^cn^d\\
\Psi_4&=&C_{abcd}\bar{m}^an^b\bar{m}^cn^d,
\end{eqnarray}
where $C_{abcd}$ is the Weyl tensor. For the Schwarzschild-de Sitter spacetime, the dyad components $\Psi_0$, $\Psi_1$, $\Psi_3$, $\Psi_4$ are zero by inspection of the spin coefficient equations involving them. (The list of spin coefficient equations are found in the appendix.) For $\Psi_2$, consider the spin coefficient equation $D\rho'=\rho'\rho-\Psi_2-\Lambda/3$, where
\begin{eqnarray}
D\rho'&=&-\frac{\Lambda}{6}-\frac{1}{2}r^{-2}+2Mr^{-3}\\
\rho'\rho&=&\frac{\Lambda}{6}-\frac{1}{2}r^{-2}+Mr^{-3}.
\end{eqnarray}
Then we have $\Psi_2=-Mr^{-3}$, i.e. the Schwarzschild-de Sitter mass is $M=-\Psi^o_2$, where $\Psi_2=\Psi^o_2r^{-3}$.

\section{Asymptotically de Sitter spacetimes}

The generalisation to asymptotically de Sitter spacetimes would be to have a null tetrad of the form:
\begin{eqnarray}
\vec{l}&=&\vec{\partial}_r\\
\vec{n}&=&\vec{\partial}_u+U\vec{\partial}_r+X^\theta\vec{\partial}_\theta+X^\phi\vec{\partial}_\phi\\
\vec{m}&=&\omega\vec{\partial}_r+\xi^\theta\vec{\partial}_\theta+\xi^\phi\vec{\partial}_\phi\\
\vec{\bar{m}}&=&\bar{\omega}\vec{\partial}_r+\overline{\xi^\theta}\vec{\partial}_\theta+\overline{\xi^\phi}\vec{\partial}_\phi,
\end{eqnarray}
where
\begin{eqnarray}
U&=&\frac{\Lambda}{6}r^2+U^o_0+O(r^{-1})\\
X^\mu&=&(X^\mu)^o_1r^{-1}+O(r^{-2})\\
\omega&=&\omega^o_1r^{-1}+O(r^{-2})\\
\xi^\mu&=&(\xi^\mu)^o_1r^{-1}+O(r^{-2}).
\end{eqnarray}
In writing this, we adopt the notation that a function $f(u,r,\theta,\phi)$ can be expanded as a series $\displaystyle f(u,r,\theta,\phi)=\sum_i^N{f^o_{i}(u,\theta,\phi)r^{-i}}$ in $r$ for sufficiently large $N$, and the superscript $^o$ means that the coefficients $f^o_i$ are independent of $r$. Note in our generalisation that $U$ has no term of order $r$, since the mass only appears in the order of $r^{-1}$ for the Schwarzschild-de Sitter spacetime \footnote{By studying the conformally rescaled de Sitter-like spacetime, Szabados and Tod showed that one can set the term of order $r$ for $U$ (denoted by $U^o_{-1}$) to zero using a gauge freedom arising from $(\nabla_a\Omega)(\nabla^a\Omega)=\Lambda/3+O(\Omega^2)$, where $\Omega$ is the conformal factor \cite{Szabados}.}. Similar to the asymptotically flat case \cite{newpen62,newunti62}, such a setup corresponds to the spacetime geometry being foliated into a family of null hypersurfaces given by constant values of the $u$-coordinate, satisfying $g^{ab}u_{,a}u_{,b}=0$. Recall from the previous section that by having $\tilde{l}=\tilde{d}u$, we could define our null tetrad in Eqs. (\ref{tet1})-(\ref{tet2}), where we specified in particular, $\vec{l}=\vec{\partial}_r$. Let this null vector field $\vec{l}$ satisfy $l^a_{\ ;b}l^b=0$, so that it induces a congruence of null geodesics tangent to these null hypersurfaces of constant $u$, as well as implying that the spin coefficients $\kappa$ and $\textrm{Re}(\gamma')$ are zero. Hence in writing $\vec{l}=\vec{\partial}_r$, we have tacitly employed an affine parameter for such null geodesics as our coordinate $r$. (See sections IV and VI in Ref. \cite{newpen62}.) One is permitted to define $r$ up to a linear transformation, and we will exploit this freedom to simplify the spin coefficient $\rho$ (see below) as was also done for the asymptotically flat case \cite{newunti62}. The remaining two coordinates $\theta$ and $\phi$ here denote general coordinates (so they do not necessarily represent the usual spherical coordinates) that would serve as labels for these null geodesics on each null hypersurface of constant $u$. Incidentally, whilst one may choose a different family of null hypersurfaces arising from another $u'$ coordinate and obtain a different null tetrad (specifically, $\tilde{l'}=\tilde{d}u'$ would be different), the same results would be produced since the physical outcome is independent of the frames and coordinates. (See section IV of Ref. \cite{newunti62}.)

Having determined $\vec{l}$, the first of our null tetrad vectors, there is freedom due to null rotation of the remaining three $\vec{n}$, $\vec{m}$, $\vec{\bar{m}}$ around $\vec{l}$ as well as freedom associated with spatial rotation of $\vec{m}$, $\vec{\bar{m}}$. We utilise the former to impose that $\vec{n}$, $\vec{m}$, $\vec{\bar{m}}$ are parallel transported along $\vec{l}$, leading to great simplifications since this geometric condition yields the vanishing of the spin coefficients $\gamma'$ and $\tau'$ \cite{newpen62,Pen87}. The latter freedom will be applied to set $\textrm{Im}(\gamma^o_0)=0$ for the spin coefficient $\gamma$ \cite{Szabados} (see below). Besides that, $\vec{l}=\vec{\nabla}u$ being a gradient field implies that $\rho$ is real and $\tau=\bar{\alpha}-\alpha'$ \cite{newpen62,Pen88}.

The inverse metric has components $g^{ab}=l^an^b+n^al^b-m^a\bar{m}^b-\bar{m}^am^b$, i.e.
\begin{eqnarray}
g^{ab}&=&
\begin{pmatrix}
  0 & 1 & 0 & 0 \\
  1 & 2(U-|\omega|^2) & X^\theta-2\textrm{Re}(\xi^\theta\bar{\omega}) & X^\phi-2\textrm{Re}(\xi^\phi\bar{\omega}) \\
  0 & X^\theta-2\textrm{Re}(\xi^\theta\bar{\omega}) & -2|\xi^\theta|^2 & -2\text{Re}(\xi^\theta\overline{\xi^\phi}) \\
	0 & X^\phi-2\textrm{Re}(\xi^\phi\bar{\omega}) & -2\text{Re}(\xi^\theta\overline{\xi^\phi}) & -2|\xi^\phi|^2
\end{pmatrix}\\
&=&
\begin{pmatrix}
  0 & 1 & 0 & 0 \\
  1 & \frac{\Lambda}{3}r^2+O(1) & O(r^{-1}) & O(r^{-1}) \\
  0 & O(r^{-1}) & O(r^{-2}) & O(r^{-2}) \\
	0 & O(r^{-1}) & O(r^{-2}) & O(r^{-2})
\end{pmatrix}.
\end{eqnarray}

The directional derivatives are
\begin{eqnarray}
D&=&l^a\nabla_a=\frac{\partial}{\partial r}\\
D'&=&n^a\nabla_a=\frac{\partial}{\partial u}+U\frac{\partial}{\partial r}+X^\theta\frac{\partial}{\partial\theta}+X^\phi\frac{\partial}{\partial\phi}\\
\delta&=&m^a\nabla_a=\omega\frac{\partial}{\partial r}+\xi^\theta\frac{\partial}{\partial\theta}+\xi^\phi\frac{\partial}{\partial\phi}\\
\delta'&=&\bar{m}^a\nabla_a=\bar{\omega}\frac{\partial}{\partial r}+\overline{\xi^\theta}\frac{\partial}{\partial\theta}+\overline{\xi^\phi}\frac{\partial}{\partial\phi}.
\end{eqnarray}

\newpage

The spin coefficients for asymptotically de Sitter spacetimes would have the form:
\begin{eqnarray}
\kappa&=&0, \gamma'=0, \tau'=0,\\
\kappa'&=&\kappa'^o_0+\kappa'^o_1r^{-1}+O(r^{-2})\\
\sigma&=&\sigma^o_1r^{-1}+O(r^{-2})\\
\sigma'&=&\sigma'^o_0+\sigma'^o_1r^{-1}+O(r^{-2})\\
\tau&=&\tau^o_1r^{-1}+O(r^{-2})\\
\gamma&=&\gamma^o_{-1}r+\gamma^o_2r^{-2}+O(r^{-3}), \gamma^o_{-1}\neq0\\
\rho&=&\rho^o_1r^{-1}+O(r^{-2}), \rho^o_1\neq0\\
\rho'&=&\rho'^o_{-1}r+\rho'^o_1r^{-1}+O(r^{-2}), \rho'^o_{-1}\neq0\\
\alpha&=&\alpha^o_1r^{-1}+O(r^{-2}), \alpha^o_1\neq0\\
\alpha'&=&\alpha'^o_1r^{-1}+O(r^{-2}), \alpha'^o_1\neq0.
\end{eqnarray}
The leading terms $\gamma^o_{-1}$, $\rho^o_1$, $\rho'^o_{-1}$, $\alpha^o_1$, $\alpha'^o_1$ are necessarily non-zero, due to the fact that these appear in the de Sitter spacetime as worked out in the previous section. Note in our generalisation that $\rho'$ has no term of order 1, since there is no such term for the Schwarzschild-de Sitter spacetime and the mass only appears in the order of $r^{-2}$ \footnote{Such a $\rho'$ term of order 1 (denoted by $\rho'^o_0$) would be proportional to $\rho^o_2$, due to the spin coefficient equation involving $D\rho'$. As we will set $\rho^o_2=0$ by the freedom in choosing the origin of the affine parameter $r$, this would lead to the vanishing of $\rho'^o_0$ anyway.}. Similarly, $\gamma$ has no terms of orders 1 and $r^{-1}$. We should point out however, that imposing $\textrm{Im}(\gamma^o_0)=0$ would fix the spatial orientation of $\vec{m}$ and $\vec{\bar{m}}$ \cite{Szabados} \footnote{The real part of $\gamma^o_0$ is proportional to $U^o_{-1}$ via the metric equation involving $DU$. As we have set $U^o_{-1}=0$, this would give $\textrm{Re}(\gamma^o_0)=0$. Anyway, the spin coefficient equation for $D\gamma$ would force $\gamma^o_1=0$. Thus, our stipulation of the fall-offs here having $U^o_{-1}=0$, $\rho'^o_0=0$, $\rho^o_2=0$, $\gamma^o_0=0$ and $\gamma^o_1=0$ are all merely fixing gauge freedoms (with $\gamma^o_1=0$ being true as a consequence of the Newman-Penrose equations even without making such choices) and hence do not restrict the physical generality of our solutions.}. The origin of the affine parameter $r$ is chosen so that $\rho^o_2=0$ \cite{newunti62}. We necessarily require that $\sigma^o:=\sigma^o_2\neq0$, since this would then allow for $\sigma\Psi_4$ to contribute to $\dot{\Psi}_2$, where the dot denotes derivative with respect to $u$. Observe that our specification of the asymptotic form of the spin coefficients contains the extra terms $\kappa'^o_0$ and $\sigma'^o_0$. The former would allow for $\omega^o_1\neq0$ when solving the metric equation involving $D'\omega$, without which it would lead to $\sigma^o=0$. If the latter $\sigma'^o_0=0$, then the spin coefficient equation involving $D\sigma'$ forces $\sigma^o=0$ \footnote{The fall-offs for the spin coefficients here are consistent with those in Szabados-Tod \cite{Szabados}. Note however that there, they applied a conformal rescaling which is symmetric with respect to the spin frame $o^A$ and $\iota^A$, whereas ours here would correspond to an asymmetric choice. See section 5.6 of Penrose-Rindler for details on conformal rescalings, in particular the four specific ways of distributing the conformal factor $\Omega$ to the spin frame $o^A$ and $\iota^A$, in (5.6.26) \cite{Pen87}.}.

Following the case for asymptotically flat spacetimes, we take $\Psi_0$ to have the fall-off of order $r^{-5}$, i.e. $\Psi_0=(\Psi_0)^o_5r^{-5}+O(r^{-6})$. We denote $\Psi^o_0:=(\Psi_0)^o_5$, $\Psi^1_0:=(\Psi_0)^o_6$, $\Psi^o_1=(\Psi_1)^o_4$, $\Psi^o_2=(\Psi_2)^o_3$, $\Psi^o_3=(\Psi_3)^o_2$, $\Psi^o_4=(\Psi_4)^o_1$. Furthermore, we adopt the following notation:
\begin{eqnarray}
\delta^o:=(\xi^\theta)^o\frac{\partial}{\partial\theta}+(\xi^\phi)^o\frac{\partial}{\partial\phi}, \delta'^o:=(\overline{\xi^\theta})^o\frac{\partial}{\partial\theta}+(\overline{\xi^\phi})^o\frac{\partial}{\partial\phi},
\end{eqnarray}
where $(\xi^\theta)^o:=(\xi^\theta)^o_1$, $(\xi^\phi)^o:=(\xi^\phi)^o_1$ and thus define the $\eth$ and $\eth'$ operators acting on a scalar $\eta$ with spin-weight $s$ to be \cite{Pen87}:
\begin{eqnarray}
\eth\eta=(\delta^o+2s\bar{\alpha}^o)\eta, \eth'\eta=(\delta'^o-2s\alpha^o)\eta,
\end{eqnarray}
where $\alpha^o:=\alpha^o_1$.

With the setup in place, the 38 equations of the Newman-Penrose formalism comprising the metric equations, spin coefficient equations and the Bianchi identities can be solved asymptotically away from $\mathcal{I}$. The details of these lengthy calculations are produced in the appendix, where we list the sequence in which we solve them along with what each of these equations implies for various orders of $r^{-n}$. The resulting asymptotic solutions are summarised in the next section, where we compile the null tetrad, spin coefficients, Weyl spinor, as well as the Bianchi identities.

\section{Summary}

Asymptotically de Sitter spacetimes are described by the following null tetrad:
\begin{eqnarray}
\vec{l}&=&\vec{\partial}_r\\
\vec{n}&=&\vec{\partial}_u+U\vec{\partial}_r+X^\theta\vec{\partial}_\theta+X^\phi\vec{\partial}_\phi\\
\vec{m}&=&\omega\vec{\partial}_r+\xi^\theta\vec{\partial}_\theta+\xi^\phi\vec{\partial}_\phi\\
\vec{\bar{m}}&=&\bar{\omega}\vec{\partial}_r+\overline{\xi^\theta}\vec{\partial}_\theta+\overline{\xi^\phi}\vec{\partial}_\phi,
\end{eqnarray}
where
\begin{eqnarray}
U&=&\frac{\Lambda}{6}r^2-\left(\frac{1}{2}K+\frac{\Lambda}{2}|\sigma^o|^2\right)-\textrm{Re}(\Psi^o_2)r^{-1}+O(r^{-2})\\
X^\mu&=&\frac{1}{3}\textrm{Re}(\bar{\Psi}^o_1(\xi^\mu)^o)r^{-3}+O(r^{-4})\\
\omega&=&\eth'\sigma^or^{-1}-\left(\sigma^o\eth\bar{\sigma}^o+\frac{1}{2}\Psi^o_1\right)r^{-2}+O(r^{-3})\\
\xi^\mu&=&(\xi^\mu)^or^{-1}-\sigma^o(\overline{\xi^\mu})^or^{-2}+|\sigma^o|^2(\xi^\mu)^or^{-3}+\left(\frac{1}{6}\Psi^o_0-\sigma^o|\sigma^o|^2\right)(\overline{\xi^\mu})^or^{-4}+O(r^{-5}).\nonumber\\
\end{eqnarray}
The components of the inverse metric are related to the null tetrad via $g^{ab}=l^an^b+n^al^b-m^a\bar{m}^b-\bar{m}^am^b$.

The spin coefficients are:
\begin{eqnarray}
\kappa&=&0, \gamma'=0, \tau'=0\\
\rho&=&-r^{-1}-|\sigma^o|^2r^{-3}+\left(\frac{1}{3}\textrm{Re}(\bar{\sigma}^o\Psi^o_0)-|\sigma^o|^4\right)r^{-5}+O(r^{-6})\\
\sigma&=&\sigma^or^{-2}+\left(\sigma^o|\sigma^o|^2-\frac{1}{2}\Psi^o_0\right)r^{-4}-\frac{1}{3}\Psi^1_0r^{-5}+O(r^{-6})\\
\rho'&=&-\frac{\Lambda}{6}r+\left(\frac{1}{2}K+\frac{\Lambda}{3}|\sigma^o|^2\right)r^{-1}+(\Psi^o_2+\sigma^o\dot{\bar{\sigma}}^o)r^{-2}+O(r^{-3})\\
\sigma'&=&-\frac{\Lambda}{6}\bar{\sigma}^o-\dot{\bar{\sigma}}^or^{-1}-\left(\frac{1}{2}\bar{\sigma}^oK+\frac{\Lambda}{3}\bar{\sigma}^o|\sigma^o|^2+\frac{\Lambda}{12}\bar{\Psi}^o_0\right)r^{-2}+O(r^{-3})\\
\tau&=&-\frac{1}{2}\Psi^o_1r^{-3}+O(r^{-4})\\
\gamma&=&-\frac{\Lambda}{6}r-\frac{1}{2}\Psi^o_2r^{-2}+O(r^{-3})\\
\kappa'&=&\frac{\Lambda}{3}\eth\bar{\sigma}^o+\left(\Psi^o_3-\frac{\Lambda}{12}\bar{\Psi}^o_1\right)r^{-1}+O(r^{-2})\\
\alpha&=&\alpha^or^{-1}+\bar{\alpha}^o\bar{\sigma}^or^{-2}+\alpha^o|\sigma^o|^2r^{-3}+O(r^{-4})\\
\alpha'&=&\bar{\alpha}^or^{-1}+\alpha^o\sigma^or^{-2}+\left(\bar{\alpha}^o|\sigma^o|^2+\frac{1}{2}\Psi^o_1\right)r^{-3}+O(r^{-4}).
\end{eqnarray}

The Weyl spinor has dyad components:
\begin{eqnarray}
\Psi_0&=&\Psi^o_0r^{-5}+\Psi^1_0r^{-6}+O(r^{-7})\\
\Psi_1&=&\Psi^o_1r^{-4}-\eth'\Psi^o_0r^{-5}+O(r^{-6})\\
\Psi_2&=&\Psi^o_2r^{-3}-\left(\eth'\Psi^o_1-\frac{\Lambda}{6}\bar{\sigma}^o\Psi^o_0\right)r^{-4}+O(r^{-5})\\
\Psi_3&=&\Psi^o_3r^{-2}-\left(\eth'\Psi^o_2-\frac{\Lambda}{3}\bar{\sigma}^o\Psi^o_1\right)r^{-3}+O(r^{-4})\\
\Psi_4&=&\Psi^o_4r^{-1}-\left(\eth'\Psi^o_3-\frac{\Lambda}{2}\bar{\sigma}^o\Psi^o_2\right)r^{-2}+O(r^{-3}),
\end{eqnarray}
with the Bianchi identities:
\begin{eqnarray}
\dot{\Psi}^o_0&=&\eth\Psi^o_1+3\sigma^o\Psi^o_2+\frac{\Lambda}{6}\Psi^1_0\label{Biazero}\\
\dot{\Psi}^o_1&=&\eth\Psi^o_2+2\sigma^o\Psi^o_3-\frac{\Lambda}{6}\eth'\Psi^o_0\\
-\frac{\partial}{\partial u}(\Psi^o_2+\sigma^o\dot{\bar{\sigma}}^o)&=&-|\dot{\sigma}^o|^2-\eth\Psi^o_3-\frac{\Lambda}{3}\left(K|\sigma^o|^2-\sigma^o\eth'\eth\bar{\sigma}^o\right)+\frac{\Lambda}{6}\eth'\Psi^o_1-\frac{2\Lambda^2}{9}|\sigma^o|^4-\frac{\Lambda^2}{18}\textrm{Re}(\bar{\sigma}^o\Psi^o_0).\nonumber\\\label{Biatwo}
\end{eqnarray}

Also:
\begin{eqnarray}
&&\textrm{Im}(\Psi^o_2+\sigma^o\dot{\bar{\sigma}}^o)=\textrm{Im}(\eth'^2\sigma^o)\\
&&\Psi^o_3=-\eth\dot{\bar{\sigma}}^o-\frac{\Lambda}{6}\bar{\Psi}^o_1+\eth'U^o_0+\frac{\Lambda}{3}\sigma^o\eth'\bar{\sigma}^o,\\
&&\textrm{where\ }U^o_0=-\frac{1}{2}K-\frac{\Lambda}{2}|\sigma^o|^2\\
&&\Psi^o_4=-\ddot{\bar{\sigma}}^o+\frac{\Lambda}{3}\left(K\bar{\sigma}^o-\eth'\eth\bar{\sigma}^o\right)+\frac{2\Lambda^2}{9}\bar{\sigma}^o|\sigma^o|^2+\frac{\Lambda^2}{36}\bar{\Psi}^o_0\label{psi4}\\
&&K=1+\frac{2\Lambda}{3}\int{\text{Re}(\eth^2\bar{\sigma}^o)du},\label{K}
\end{eqnarray}
where $K$ is the Gauss curvature of the compact 2-surface of constant $u$ on $\mathcal{I}$.

Note that when $\textrm{Re}(\eth^2\bar{\sigma}^o)\neq0$, (34) (Eq (\ref{Gauss})) implies that the Gauss curvature of the 2-surfaces on $\mathcal{I}$ for fixed values of $u$ is not a constant but depends on $u$ (note that $\Theta=1/2$) \footnote{Two times the real part of $-\rho\rho'+\sigma\sigma'-\Psi_2+\Lambda/6$ has the interpretation of being the Gauss curvature of the 2-surfaces of constant $u$ on $\mathcal{I}$. See (4.14.21) in Ref. \cite{Pen87}.}. The angular coordinates $\theta$ and $\phi$ would then not correspond to those spherical coordinate angles as used for the unit sphere, but instead denote generalised coordinates intrinsic to these 2-surfaces. Furthermore, the pair of equations for $(\dot{\xi}^\theta)^o$ and $(\dot{\xi}^\phi)^o$ in Eq. (\ref{xitheta}) imply that when $\Lambda\neq0$, a non-zero $\sigma^o$ necessarily requires $(\xi^\theta)^o$ and $(\xi^\phi)^o$ to have a $u$-dependence. A ramification of this is that $\mathcal{I}$ has to be non-conformally flat if $\sigma^o\neq0$, since it is no longer possible to make the usual specifications $(\xi^\theta)^o=1/\sqrt{2}$ and $(\xi^\phi)^o=i\csc{\theta}/\sqrt{2}$. This is in accordance to what has been pointed out by Ashtekar et al. by studying the conformal versions of de Sitter-like spacetimes \cite{ash1}.

In Section 6, we will illustrate this explicitly in the case of axisymmetry by solving the pair of equations Eq. (\ref{xitheta}), followed by Eq. (\ref{1}) for $\alpha^o$ (which would also satisfy Eq. (\ref{alpha})). With these expressions, we can then validate the Gauss curvature equation, given by Eq. (\ref{Gauss}), as well as showing that the Cotton tensor for the 3-manifold $\mathcal{I}$ is not zero when $\sigma^o\neq0$ --- indicating non-conformal flatness.

\section{Bondi mass-loss formula}

For asymptotically flat spacetimes, the Bondi mass is $\displaystyle M_B=-\frac{1}{A}\oint{(\Psi^o_2+\sigma^o\dot{\bar{\sigma}}^o)}d^2S$, where the integration is over a compact 2-surface of constant $u$ on $\mathcal{I}$ and $A$ is the area. A generalisation to asymptotically de Sitter spacetimes would contain corrections due to the cosmological constant $\Lambda$ \footnote{In the mass-loss formula Eq. (\ref{Bondimasslosspsi0}), the rate of mass-loss has corrections due to the cosmological constant $\Lambda$. It is hence necessary that the mass itself has corrections due to $\Lambda$. Otherwise the mass cannot possibly change differently than when $\Lambda=0$.}. From the Bianchi identity involving the $u$-derivative of $\Psi^o_2$ in Eq. (\ref{Biatwo}), we obtain:
\begin{eqnarray}
\frac{dM_B}{du}&+&\frac{1}{A}\oint{\left(\Psi^o_2+\sigma^o\dot{\bar{\sigma}}^o\right)\frac{\partial}{\partial u}(d^2S)}\nonumber\\
&=&-\frac{1}{A}\oint{\bigg(|\dot{\sigma}^o|^2+\frac{\Lambda}{3}\left(K|\sigma^o|^2+|\eth'\sigma^o|^2\right)+\frac{2\Lambda^2}{9}|\sigma^o|^4}+\frac{\Lambda^2}{18}\textrm{Re}(\bar{\sigma}^o\Psi^o_0)\bigg)d^2S,\label{Bondimasslosspsi0}
\end{eqnarray}
where the terms comprising $\eth\Psi^o_3, \eth'\Psi^o_1$ vanish upon integration over the compact 2-surface which has no boundary, and integration by parts is applied to $\sigma^o\eth'\eth\bar{\sigma}^o$. The second term on the left-hand side arises because the 2-surfaces of constant $u$ are not round 2-spheres, but generally have a $u$-dependence when $\sigma^o\neq0$.

All terms on the right-hand side of this equation have a definite sign, except for $K|\sigma^o|^2$ and $\textrm{Re}(\bar{\sigma}^o\Psi^o_0)$. If we define a new mass $M_{\Lambda}$ that includes the former (as well as the second term on the left-hand side):
\begin{eqnarray}\label{Lambdapositivemass}
M_{\Lambda}:=M_B+\frac{1}{A}\int{\left(\oint{\left(\Psi^o_2+\sigma^o\dot{\bar{\sigma}}^o\right)\frac{\partial}{\partial u}(d^2S)}\right)du}+\frac{\Lambda}{3A}\int{\left(\oint{K|\sigma^o|^2d^2S}\right)du},
\end{eqnarray}
we then have a mass-loss formula:
\begin{eqnarray}
\frac{dM_{\Lambda}}{du}
&=&-\frac{1}{A}\oint{\bigg(|\dot{\sigma}^o|^2+\frac{\Lambda}{3}|\eth'\sigma^o|^2+\frac{2\Lambda^2}{9}|\sigma^o|^4}+\frac{\Lambda^2}{18}\textrm{Re}(\bar{\sigma}^o\Psi^o_0)\bigg)d^2S.\label{0Bondimasslosspsi0}
\end{eqnarray}

In the absence of incoming radiation since we are considering an isolated system, $\Psi^o_0=0$ \footnote{Eq. (\ref{Bondimasslosspsi0}) is concerned with taking the closed integral over a compact 2-surface of constant $u$ on $\mathcal{I}$. In order to eradicate its presence entirely from the mass-loss formula, this necessitates $\Psi^o_0(u,\theta,\phi)=0$ for all values of $u, \theta, \phi$ on $\mathcal{I}$ where $r\rightarrow\infty$. But since $\Psi^o_0(u,\theta,\phi)$ does not depend on $r$, this implies that it is zero for all values of $r$, i.e. it is zero everywhere. See also the Discussion Section and Fig. \ref{fig1} there.}, so:
\begin{eqnarray}\label{Bondimasslosspositive}
\frac{dM_{\Lambda}}{du}
&=&-\frac{1}{A}\oint{\bigg(|\dot{\sigma}^o|^2+\frac{\Lambda}{3}|\eth'\sigma^o|^2+\frac{2\Lambda^2}{9}|\sigma^o|^4}\bigg)d^2S.
\end{eqnarray}

Remarks:
\begin{enumerate}
	\item When $\Lambda=0$, the null tetrad, spin coefficients, Weyl spinor, Bianchi identities, mass-loss formula, all reduce exactly to the known case of asymptotically flat spacetimes \footnote{See Penrose-Rindler \cite{Pen88} for the asymptotically flat vacuum solutions, though they also included the Maxwell source there and they used complex stereographic angles for the 2-spheres. Note the typoes there: a missing bar on $\alpha^0$ for $\alpha$ for order $r^{-2}$, a missing factor of 1/2 on $\sigma'$ for order $r^{-2}$ as well as a missing $^0$ on $\dot{\bar{\sigma}}$ for $\sigma'$ for order $r^{-3}$, and their $\alpha^0$ should be $\bar{\zeta}/2\sqrt{2}$. There is also a misprint on $\xi^3$.}.
	\item In the absence of incoming radiation for $\Lambda>0$, i.e. asymptotically de Sitter spacetimes, positivity of the mass-loss of $M_\Lambda$ is manifest \footnote{For clarity, the negative sign on the right-hand side of Eq. (\ref{Bondimasslosspositive}) implies that the mass $M_{\Lambda}$ decreases as $u$ approaches the source's timelike infinity $i^+$. So, we say that ``the mass-loss is positive'' to mean that the mass $M_{\Lambda}$ experiences a decrease.}. This is perhaps the simplest generalisation of the Bondi mass such that the mass-loss of $M_{\Lambda}$ is strictly positive.
	\item This entire mathematics also applies for $\Lambda<0$, i.e. asymptotically anti-de Sitter spacetimes. However, positivity of the mass-loss is not guaranteed since the term with $\Lambda$ is of the opposite sign. Nevertheless, one may consider defining the generalised mass to include the $u$-integral of that term, and obtain a strictly positive mass-loss formula.
	\item As pointed out near the end of the previous section, a conformally flat $\mathcal{I}$ when $\Lambda\neq0$ implies that $\sigma^o=0$. So gravitational waves may only carry energy away from the source with a non-conformally flat $\mathcal{I}$.
	\item When $\Lambda\neq0$, the mass-loss can be non-zero even if $\sigma^o$ does not depend on $u$. In fact, $\dot{\sigma}^o$ only contributes to the mass-loss in the single term not involving $\Lambda$. This may also be seen in the dyad component of the Weyl spinor $\Psi^o_4$ in Eq. (\ref{psi4}), where $\ddot{\bar{\sigma}}^o$ appears in the single term not involving $\Lambda$, with $\sigma^o$ itself contributing due to the non-zero $\Lambda$.
\end{enumerate}

\section{The 2-surfaces of constant $u$ on $\mathcal{I}$, axisymmetry}

We now consider the pair of equations in Eq. (\ref{xitheta}) for the 2-surfaces of constant $u$ on $\mathcal{I}$, where the metric is $g_2=\vec{m}\otimes\vec{\bar{m}}+\vec{\bar{m}}\otimes\vec{m}$, with $\vec{m}=(\xi^\theta)^o\vec{\partial}_\theta+(\xi^\phi)^o\vec{\partial}_\phi$. In general, gravitational waves have two polarisations or radiative modes \cite{ash2} which are here encoded into the real and imaginary parts of $\sigma^o$. Collecting real and imaginary parts of Eq. (\ref{xitheta}) would give a set of coupled differential equations for $(\xi^\theta)^o$ and another corresponding set for $(\xi^\phi)^o$. It does not appear however, that these sets of coupled differential equations may be analytically solved for $(\xi^\theta)^o$ and $(\xi^\phi)^o$ in terms of some general real and imaginary parts of $\sigma^o$. Nevertheless, it is known from the case of axisymmetry \cite{chi1} where the metric components are explicitly independent of $\phi$, that the emitted gravitational waves only possess a single radiative mode \cite{ash4}. We shall hence proceed here by setting say $\textrm{Im}(\sigma^o)=0$, as well as imposing $\phi$-independence to solve the pair of equations Eq. (\ref{xitheta}) for an axisymmetric isolated gravitating system. In that case, we find that
\begin{eqnarray}
(\xi^\theta)^o&=&\frac{1}{\sqrt{2}}e^{-\Lambda f(u,\theta)}\label{xisolone}\\
(\xi^\phi)^o&=&\frac{1}{\sqrt{2}}ie^{\Lambda f(u,\theta)}\csc{\theta},\label{xisoltwo}
\end{eqnarray}
where $3f(u,\theta)=\int{\sigma^o(u,\theta)du}$. With this, the metric for the 2-surface of constant $u$ on $\mathcal{I}$ is
\begin{eqnarray}\label{2sur}
g_2=e^{2\Lambda f(u,\theta)}d\theta^2+e^{-2\Lambda f(u,\theta)}\sin^2{\theta}d\phi^2.
\end{eqnarray}
Subsequently, the first of Eq. (\ref{1}) says that $\alpha^o$ is real, with the second one yielding
\begin{eqnarray}
\alpha^o=(\xi^\theta)^o\left(-\frac{1}{2}\cot{\theta}+\frac{\Lambda}{2}\frac{\partial f}{\partial\theta}\right).
\end{eqnarray}
One can verify that Eq. (\ref{alpha}) is automatically satisfied for this $\alpha^o$. The Gauss curvature for this 2-surface in Eq. (\ref{2sur}) may be calculated using $K=-(G_\theta/\sqrt{EG})_\theta/2\sqrt{EG}$, where $E=e^{2\Lambda f(u,\theta)/3}$ and $G=e^{-2\Lambda f(u,\theta)/3}\sin^2{\theta}$, to get $K=4\textrm{Re}(\delta^o\alpha^o)-8|\alpha^o|^2$. Taking the $u$-derivative of $2\textrm{Re}(\delta^o\alpha^o)-4|\alpha^o|^2$, one would find that it is equal to $\Lambda\eth^2\bar{\sigma}^o/3$, hence corroborating with the Gauss curvature equation of Eq. (\ref{Gauss}) \cite{Pen87} \footnote{In doing these calculations, although $\sigma^o$ here is real, one should remember that what appears is $\bar{\sigma}^o$, i.e. a quantity with spin-weight $-2$, and apply the definitions of $\eth$ and $\eth'$ appropriately. Keeping track of whether a real function is actually referring to its complex conjugate for spin-weighted quantities is thus vital, in order to correctly relate to the $\eth$ and $\eth'$ operators.}. Explicitly, the expression for the Gauss curvature $K$ is:
\begin{eqnarray}\label{axiGauss}
K=e^{-2\Lambda f}\left(1+3\Lambda\frac{\partial f}{\partial\theta}\cot{\theta}+\Lambda\frac{\partial^2f}{\partial\theta^2}-2\Lambda^2\left(\frac{\partial f}{\partial\theta}\right)^2\right).
\end{eqnarray}
When $\sigma^o=0$, then $(\xi^\theta)^o=1/\sqrt{2}$, $(\xi^\phi)^o=i\csc{\theta}/\sqrt{2}$, $\alpha^o=-\cot{\theta}/2\sqrt{2}$, and $K=1$, as anticipated for de Sitter spacetime.

Furthermore, one can then obtain the 3-metric of the axisymmetric conformally rescaled $\mathcal{I}$ (i.e. apply conformal rescaling to the full axisymmetric (3+1)-d spacetime with conformal factor $\Omega=1/r+O(1/r^2)$, and take the limit $r\rightarrow\infty$):
\begin{eqnarray}\label{scri}
g_{\mathcal{I}}=-\frac{\Lambda}{3}du^2-e^{2\Lambda f(u,\theta)/3}d\theta^2-e^{-2\Lambda f(u,\theta)/3}\sin^2{\theta}d\phi^2.
\end{eqnarray}
The Cotton tensor for this 3-metric may be calculated and it generally does not vanish unless $f$ is constant, i.e. $\sigma^o=0$. It is thus non-conformally flat when outgoing radiation depletes the energy of the isolated source. This is in fact, the form of the axisymmetric $\mathcal{I}$ studied by He and Cao \cite{chi1}.

Incidentally, direct calculation gives $\displaystyle\eth'\sigma^o=\frac{1}{\sqrt{2}}e^{-\Lambda f}\left(\frac{\partial\sigma^o}{\partial\theta}+2\sigma^o\cot{\theta}-2\Lambda\sigma^o\frac{\partial f}{\partial\theta}\right)$, so the mass-loss formula Eq. (\ref{Bondimasslosspositive}) is:
\begin{eqnarray}\label{masslossaxi}
\frac{dM_\Lambda}{du}=-\frac{1}{A}\oint{\left(|\dot{\sigma}^o|^2+\frac{\Lambda}{6}e^{-2\Lambda f}\left(\frac{\partial\sigma^o}{\partial\theta}+2\sigma^o\cot{\theta}-2\Lambda\sigma^o\frac{\partial f}{\partial\theta}\right)^2+\frac{2\Lambda^2}{9}|\sigma^o|^4\right)d^2S}.\ 
\end{eqnarray}

\section{Discussion}

By extending Newman-Unti's approach in solving the vacuum Newman-Penrose equations with non-zero $\Lambda$ asymptotically, we have been able to generalise the Bondi mass (Eq. (\ref{Lambdapositivemass})) and obtain the mass-loss formula (Eq. (\ref{Bondimasslosspositive}) for no incoming radiation) for an isolated system emitting gravitational waves in a universe with $\Lambda>0$. Our use of spherical coordinates implies that this holds also for $\Lambda<0$, and $\Lambda=0$ reduces to the asymptotically flat vacuum spacetimes, despite the distinct properties of $\mathcal{I}$ being spacelike, timelike and null for de Sitter, anti-de Sitter and Minkowski-like spacetimes, respectively. This reduction is perhaps not a surprise, since in a way, Penrose's conclusions of $\mathcal{I}$ being spacelike, timelike or null may be deduced from the use of spherical coordinates in Eq. (\ref{scri}) \footnote{Whilst Eq. (\ref{scri}) is for axisymmetry, the general case with both radiative modes would be described by a complex $\sigma^o$. Consequently, the solutions for $(\xi^\theta)^o$ and $(\xi^\phi)^o$ in Eqs. (\ref{xisolone}) and (\ref{xisoltwo}) are both complex functions of the complex $\sigma^o$ (unlike the case with just one radiative mode where $\sigma^o$ is a real function, such that one may choose $(\xi^\theta)^o$ to be real and $(\xi^\phi)^o$ to be imaginary, as is done in Eqs. (\ref{xisolone}) and (\ref{xisoltwo})), and the metric $g_2$ for the 2-surfaces of constant $u$ on $\mathcal{I}$ in Eq. (\ref{2sur}) would not be diagonal.}.

Notice how a non-zero $\Lambda$ introduces new physics via the Newman-Penrose equations. Well, Eqs. (\ref{sigmaprime}) and (\ref{kappaprime}) (together with several other equations) respectively dictate that the spin coefficients $\sigma'=-\Lambda\bar{\sigma}^o/6+O(r^{-1})$ and $\kappa'=\Lambda\eth\bar{\sigma}^o/3+O(r^{-1})$ are generally non-vanishing on $\mathcal{I}$. These are consequences of $\mathcal{I}$ being non-null together with a non-zero $\sigma^o$. (See also Ref. \cite{Szabados} where they derived relations involving $\sigma'+\bar{\sigma}$ as well as for $\kappa'$ which relate to a spacelike $\mathcal{I}$, using the conformally rescaled de Sitter-like spacetime. But take note of their conformal rescaling which is symmetric with respect to the spin frame $o^A$ and $\iota^A$, whereas ours here would correspond to an asymmetric choice.) Apart from that, Eq. (\ref{xitheta}) demands that conformal flatness of $\mathcal{I}$ implies $\sigma^o=0$, such that $\Psi^o_4$ which represents outgoing radiation would essentially vanish. The non-conformal flatness of $\mathcal{I}$ when $\sigma^o\neq0$ also shows up in Eq. (\ref{Gauss}) for the Gauss curvature of the 2-surfaces of constant $u$ on $\mathcal{I}$ which is no longer a constant, but picks up a term involving $2\Lambda\int{\textrm{Re}(\eth^2\bar{\sigma}^o)du}/3$. This is illustrated explicitly for the case of axisymmetry, where the Cotton tensor of $\mathcal{I}$ is generally non-zero when $\sigma^o\neq0$. As a result, the mass-loss formula Eq. (\ref{Bondimasslosspsi0}) has corrections due to the cosmological constant.

\begin{figure}
\centering
\includegraphics[width=15cm]{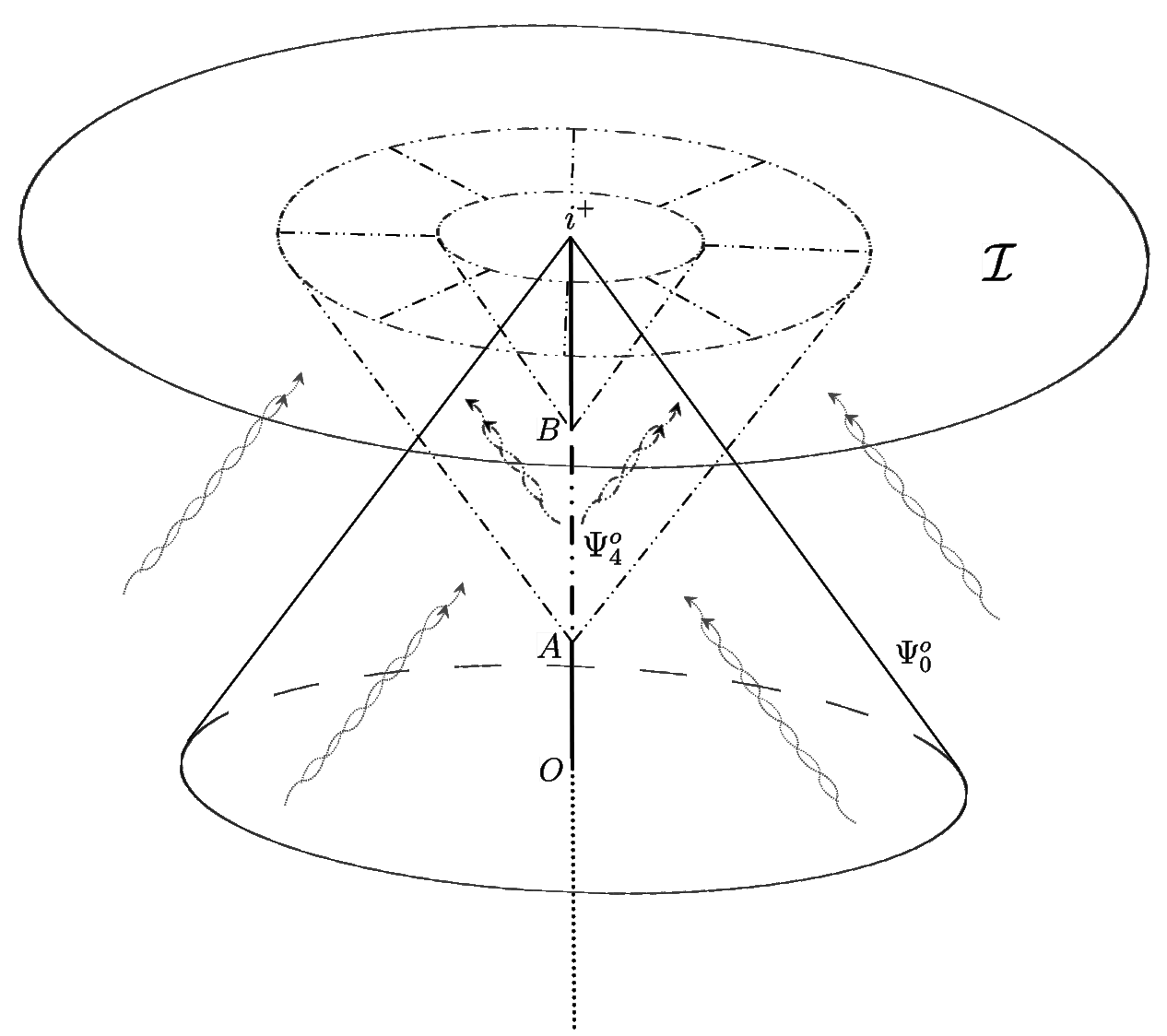}
\caption{A scenario of an isolated gravitating system in a universe with $\Lambda>0$ beginning at some time $t$, represented by point $O$ in the Penrose diagram (with one spatial dimension suppressed). In this example, the isolated system only emits gravitational radiation between the time interval represented by the points $A$ and $B$. The outgoing radiation $\Psi^o_4$ carries energy away from the isolated system, eventually arriving at the spacelike $\mathcal{I}$. If incoming radiation $\Psi^o_0$ is present, those beyond the isolated system's cosmological horizon (which are causally disconnected from it) also end up at $\mathcal{I}$, and would get picked up by the mass-loss formula Eq. (\ref{Bondimasslosspsi0}) when $\sigma^o\neq0$ --- more specifically when $\textrm{Re}(\bar{\sigma}^o\Psi^o_0)\neq0$. No incoming radiation corresponds to $\Psi^o_0=0$ everywhere.}
\label{fig1}
\end{figure}

A peculiar $\Lambda^2$-correction term involves $\Psi^o_0$, which may be interpreted as incoming radiation (i.e. heading towards the isolated system). As explained by Ashtekar et al. for $\Lambda>0$ \cite{ash2,ash4}, the appearance of negative energy carried by outgoing gravitational waves are associated with the time translation Killing vector becoming spacelike and ``past directed'' beyond the cosmological horizon of the isolated source. The outgoing radiation from the source $\Psi^o_4$ that reaches $\mathcal{I}$ is distinct from the incoming radiation $\Psi^o_0$ from beyond the source's cosmological horizon which also reaches $\mathcal{I}$, but cannot influence the source (see Fig. \ref{fig1}). (This does not occur in the asymptotically flat case, i.e. the incoming radiation does not reach $\mathcal{I}^+$. There, $\mathcal{I}^+$ is a null hypersurface ``parallel'' to the incoming radiation \cite{Haw}.) Due to the Bianchi identity Eq. (\ref{Biazero}), one is only permitted to specify the function $\Psi^o_0(u,\theta,\phi)$ on some initial hypersurface $u=u_0$, such that its values at the next hypersurface $u_0\rightarrow u_0+du$ is determined from this differential equation for $\dot{\Psi}^o_0$. Imposing $\Psi^o_0=0$ for all values of $u$ instead of just the single initial hypersurface $u=u_0$ gives a relation between $\eth\Psi^o_1$, $3\sigma^o\Psi^o_2$ and $\Lambda\Psi^1_0/6$ (where recall that $\Psi^1_0:=(\Psi_0)^o_6$, i.e. the next order $r^{-6}$ of $\Psi_0$). If $\Lambda=0$, this would be a constraint between $\eth\Psi^o_1$ and $3\sigma^o\Psi^o_2$. Remarkably (or perhaps fortuitously), this is not so for $\Lambda\neq0$, since there is here the free function $\Lambda\Psi^1_0/6$ that would now play the role of $\dot{\Psi}^o_0$, had $\Psi^o_0$ not been set to zero. The same Bianchi identity for the next order would give $\dot{\Psi}^1_0$ involving $\Lambda\Psi^2_0$ (where $\Psi^2_0:=(\Psi_0)^o_7$), and so on. Ergo, whilst incoming radiation in the asymptotically flat case corresponds to setting $\Psi^o_0=0$ on some initial null hypersurface $u=u_0$, the situation with $\Lambda\neq0$ would be to set $\Psi^o_0=0$ everywhere.

Our mass-loss formula Eq. (\ref{Bondimasslosspositive}) and the proposed generalised Bondi mass Eq. (\ref{Lambdapositivemass}) affirms a fundamental physical property first shown by Bondi in 1962: \emph{In the absence of incoming radiation ($\Psi^o_0=0$), the mass of an isolated gravitating system strictly decreases when it radiates gravitational waves, with the latter carrying energy away}. Overall, the correction terms are due to the cosmological constant giving rise to a background curvature even in vacuum/empty space. If we possess extremely high precision tools, in principle, measurements would reveal deviations from the previous theory with $\Lambda=0$, and our work here provides a framework upon which the value of $\Lambda$ may be independently determined from experiments/observations/measurements related to gravitational waves. Such corrections however, are negligible for astrophysical sources that ongoing detectors are focused on \cite{ash3,ash4}. (Also, see section 4.1 (c) of Ref. \cite{CWill} on a post-Newtonian study.)

A linearised expression for the energy carried away by gravitational waves has been derived by Ashtekar et al. \cite{ash3,ash4} (see Eq. (6) of Ref. \cite{ash4}). Note however that they based their study upon a form of the de Sitter metric which is directly related to $-dt^2+e^{2Ht}(dx^2+dy^2+dz^2)$ as well as a Killing vector generating time-translation that has a factor of $H$ (see Eqs. (2.1), (2.2) (2.3), (2.4) in Ref. \cite{ash3}), where $H=\sqrt{\Lambda/3}$. Several other quantities involving $H$ are subsequently used, hence their linearised expression for the energy carried away by gravitational waves would involve half-integer powers of $\Lambda$. In contrast, our work here uses spherical coordinates (Eq. (\ref{deSittermetric})) with no square root of $\Lambda$ involved. Furthermore, no square root of $\Lambda$ ever appears throughout the solving process of the 38 Newman-Penrose equations. (Almost all these differential equations are linear.) It is shown explicitly in the case of axisymmetry that the mass-loss formula Eq. (\ref{masslossaxi}) contains a factor of $e^{-2\Lambda f}$, arising from $|\eth'\sigma^o|^2$. The Gauss curvature $K$ (which is involved in the generalised mass $M_\Lambda$ in Eq. (\ref{Lambdapositivemass})) also has this factor $e^{-2\Lambda f}$, as worked out in Eq. (\ref{axiGauss}). If a series expansion is carried out, then there would be terms involving all positive integer powers of $\Lambda$. An approximation would only produce a subset containing some of the lower order terms. 

Other than the Bondi mass, there is another widely used quantity first given by Arnowitt, Deser and Misner \cite{adm} for zero cosmological constant, referred to as the \emph{ADM mass} \cite{adm,Poisson}. The ADM mass differs from the Bondi mass of an isolated gravitating system that emits gravitational waves. Whilst the latter decreases due to energy carried away by gravitational waves, the former remains constant. Nevertheless, these two notions of mass-energy are equivalent for stationary spacetimes. A good description on these two masses may be found in Ref. \cite{Poisson}: After deriving a general expression for the Hamiltonian of a general spacetime as an integral over a 2-surface (for zero cosmological constant), the ADM mass is obtained by making a choice of the lapse and shift, then taking the limit where the 2-surface goes to spatial infinity, whilst the Bondi mass is arrived at by taking the limit to null infinity.

We are not aware of any existing way of defining the ADM mass for asymptotically flat spacetimes, using the Newman-Unti approach \cite{newunti62}. Essentially, the setup here is made with the purpose of going to null infinity, instead of spatial infinity. Nevertheless for stationary spacetimes, the Bondi mass (for $\Lambda=0$) reduces to $\displaystyle M_B=\oint{\Psi^o_2\ d^2S}$, where $\Psi^o_2$ is $u$-independent. This expression for the Bondi mass of a stationary spacetime must also be equal to its ADM mass. For a general spacetime (for $\Lambda=0$), one can move the right-hand term of the mass-loss formula in Eq. (\ref{one}) to the inside of the $u$-derivative on the left-hand side and define the quantity $\displaystyle Q=-\frac{1}{A}\oint{\left(\Psi^o_2+\int{\sigma^o\ddot{\bar{\sigma}}^odu}\right)d^2S}$. The mass-loss formula Eq. (\ref{one}) then says that this $Q$ is a constant, i.e. $dQ/du=0$ \footnote{It would then be a good exercise to show that this is equivalent/not equivalent to any of the standard ADM mass definitions.}. One can also do this for $\Lambda\neq0$ with our mass-loss formula Eq. (\ref{Bondimasslosspsi0}) to define the corresponding $Q_\Lambda$ such that $dQ_\Lambda/du=0$, and this would be a conserved quantity. There are approaches that relate the ADM and Bondi masses involving the covariant phase space formalism, viz. the work by Ashtekar, Bombelli and Reula \cite{200years}, with Wald and Zoupas \cite{WaldZoupas} using a symplectic potential to produce general conserved quantities when there does not exist a Hamiltonian generating the asymptotic symmetry. These \cite{200years,WaldZoupas} (as well as \cite{Poisson}) provide a unified treatment of both the ADM and Bondi masses for asymptotically flat spacetimes, and it is certainly intriguing to employ the framework of Ref. \cite{WaldZoupas} in attempting to define the mass of a spacetime with $\Lambda\neq0$ via a symplectic potential.

Extensions to couple with Maxwell fields in a similar way should be straightforward, and will be presented elsewhere \cite{Vee2016b}. With these electro-$\Lambda$ asymptotic solutions, one can generalise some known results for asymptotically flat spacetimes (like say those in Refs. \cite{framemory,Newmanenigma}) to include the cosmological constant. Also, whilst we have assumed that the quantities may be expanded as a series with sufficiently many orders away from $\mathcal{I}$, one may follow Newman-Penrose's original study \cite{newpen62} to make minimal assumptions on the differentiability criteria, and sedulously work out the detailed consequences.

\newpage

\appendix*

\section{Details on solving the 38 equations of the Newman-Penrose formalism (metric equations, spin coefficient equations, Bianchi identities), order-by-order away from $\mathcal{I}$}

The metric equations are
\begin{eqnarray}
{[}\delta,D]r &:& (9)\ D\omega=\rho\omega+\sigma\bar{\omega}+\alpha'-\bar{\alpha}\\
{[}D',D]r &:& (14)\ DU=2\textrm{Re}(\bar{\tau}\omega-\gamma)\\
{[}\delta,D]\theta &:& (16)\ D\xi^\theta=\rho\xi^\theta+\sigma\overline{\xi^\theta}\\
{[}\delta,D]\phi &:& (17)\ D\xi^\phi=\rho\xi^\phi+\sigma\overline{\xi^\phi}\\
{[}D',D]\theta &:& (18)\ DX^\theta=2\textrm{Re}(\bar{\tau}\xi^\theta)\\
{[}D',D]\phi &:& (19)\ DX^\phi=2\textrm{Re}(\bar{\tau}\xi^\phi)\\
{[}\delta,D']r &:& (20)\ D'\omega-\delta U=(\rho'+2i\textrm{Im}(\gamma))\omega+\bar{\sigma}'\bar{\omega}-\bar{\kappa}'\\
{[}\delta',\delta]r &:& (21)\ \textrm{Im}(\delta'\omega)=\textrm{Im}(\rho'+(\alpha+\bar{\alpha}')\omega)\\
{[}\delta,D']\theta &:& (26)\ D'\xi^\theta-\delta X^\theta=(\rho'+2i\textrm{Im}(\gamma))\xi^\theta+\bar{\sigma}'\overline{\xi^\theta}\\
{[}\delta,D']\phi &:& (27)\ D'\xi^\phi-\delta X^\phi=(\rho'+2i\textrm{Im}(\gamma))\xi^\phi+\bar{\sigma}'\overline{\xi^\phi}\\
{[}\delta',\delta]\theta &:& (32)\ \textrm{Im}(\delta'\xi^\theta)=\textrm{Im}((\alpha+\bar{\alpha}')\xi^\theta)\\
{[}\delta',\delta]\phi &:& (33)\ \textrm{Im}(\delta'\xi^\phi)=\textrm{Im}((\alpha+\bar{\alpha}')\xi^\phi).
\end{eqnarray}
\newpage

The spin coefficient equations are
\begin{eqnarray}
(1)\ D\sigma'&=&\sigma'\rho+\rho'\bar{\sigma}\\
(2)\ D\rho&=&\rho^2+\sigma\bar{\sigma}\\
(3)\ D\sigma&=&2\rho\sigma+\Psi_0\\
(8)\ D\rho'&=&\rho'\rho+\sigma'\sigma-\Psi_2-\frac{\Lambda}{3}\\
(10)\ D\alpha&=&\alpha\rho-\alpha'\bar{\sigma}\\
(11)\ D\alpha'&=&\alpha'\rho-\alpha\sigma-\Psi_1\\
(12)\ D\tau&=&\tau\rho+\bar{\tau}\sigma+\Psi_1\ (\tau=\bar{\alpha}-\alpha')\\
(13)\ D\gamma&=&\tau\alpha-\bar{\tau}\alpha'+\Psi_2-\frac{\Lambda}{6}\\
(15)\ D\kappa'&=&\tau\sigma'+\bar{\tau}\rho'-\Psi_3\\
(22)\ D'\sigma-\delta\tau&=&\rho\bar{\sigma}'+\sigma(\rho'+2\gamma+2i\textrm{Im}(\gamma))+2\alpha'\tau\\
(23)\ D'\rho-\delta'\tau&=&\sigma\sigma'+\rho(\bar{\rho}'+2\textrm{Re}(\gamma))-2\alpha\tau-\Psi_2-\frac{\Lambda}{3}\\
(24)\ D'\rho'-\delta\kappa'&=&\sigma'\bar{\sigma}'+\rho'(\rho'-2\textrm{Re}(\gamma))-2\alpha'\kappa'\\
(25)\ D'\sigma'-\delta'\kappa'&=&\sigma'(2\textrm{Re}(\rho')-2\gamma-2i\textrm{Im}(\gamma))+2\alpha\kappa'+\Psi_4\\
(28)\ D'\alpha-\delta'\gamma&=&\alpha(\bar{\rho}'-2i\textrm{Im}(\gamma))+(\bar{\alpha}-2\alpha')\sigma'-\rho\kappa'-\Psi_3\\
(29)\ D'\alpha'+\delta\gamma&=&\alpha'(\rho'+2i\textrm{Im}(\gamma))-\alpha\bar{\sigma}'+\sigma\kappa'-\tau\rho'\\
(30)\ \delta\rho-\delta'\sigma&=&(\bar{\alpha}-\alpha')\rho-(3\alpha+\bar{\alpha}')\sigma-\Psi_1\\
(31)\ \delta'\rho'-\delta\sigma'&=&(\bar{\alpha}'-\alpha)\rho'-(3\alpha'+\bar{\alpha})\sigma'-\Psi_3\\
(34)\ \delta\alpha+\delta'\alpha'&=&\alpha\bar{\alpha}+\alpha'\bar{\alpha}'+2\alpha\alpha'-\rho\rho'+\sigma\sigma'-\Psi_2+\frac{\Lambda}{6}.
\end{eqnarray}

The Bianchi identities are
\begin{eqnarray}
(4)\ D\Psi_1-\delta'\Psi_0&=&4\rho\Psi_1-4\alpha\Psi_0\\
(5)\ D\Psi_2-\delta'\Psi_1&=&3\rho\Psi_2-2\alpha\Psi_1+\sigma'\Psi_0\\
(6)\ D\Psi_3-\delta'\Psi_2&=&2\rho\Psi_3+2\sigma'\Psi_1\\
(7)\ D\Psi_4-\delta'\Psi_3&=&\rho\Psi_4+2\alpha\Psi_3+3\sigma'\Psi_2\\
(35)\ D'\Psi_0-\delta\Psi_1&=&(4\gamma+\rho')\Psi_0-2(2\tau-\alpha')\Psi_1+3\sigma\Psi_2\\
(36)\ D'\Psi_1-\delta\Psi_2&=&-\kappa'\Psi_0+2(\gamma+\rho')\Psi_1-3\tau\Psi_2+2\sigma\Psi_3\\
(37)\ D'\Psi_2-\delta\Psi_3&=&-2\kappa'\Psi_1+3\rho'\Psi_2-2(\tau+\alpha')\Psi_3+\sigma\Psi_4\\
(38)\ D'\Psi_3-\delta\Psi_4&=&-3\kappa'\Psi_2-2(\gamma-2\rho')\Psi_3-(\tau+4\alpha')\Psi_4.
\end{eqnarray}

The metric equations, spin coefficient equations, and the Bianchi identities are solved for various orders beginning from the highest non-trivial power of $r$, in the following sequence as indicated by the number in front of the equation. Listed below are the results when solving them order-by-order. We begin by solving the radial equations having the $D=\partial/\partial r$ derivative. Well, ``$0=0$'' means that there is no information or no \emph{new} information at this order, i.e. the equation is identically satisfied at this order.

(1) $D\sigma'=\sigma'\rho+\rho'\bar{\sigma}$.
\begin{eqnarray}
1&:&\sigma^o_1=0\\
r^{-1}&:&\sigma'^o_0=-\frac{\rho'^o_{-1}}{\rho^o_1}\bar{\sigma}^o.\\
&\ &\textrm{So }\sigma'^o_0=-\frac{\Lambda}{6}\bar{\sigma}^o,\textrm{ from }\rho^o_1=-1\textrm{ in (2), }\rho'^o_{-1}=-\frac{\Lambda}{6}\textrm{ in (8).}\label{sigmaprime}
\end{eqnarray}

(2) $D\rho=\rho^2+\sigma\bar{\sigma}$.

(3) $D\sigma=2\rho\sigma+\Psi_0$.
\begin{eqnarray}
r^{-2}&:&
\begin{cases}
\rho^o_1(\rho^o_1+1)=0,\textrm{ so }\rho^o_1=-1, \textrm{ since }\rho^o_1\neq0\\
0=0
\end{cases}\\
r^{-3}&:&
\begin{cases}
0=0\\
0=0
\end{cases}\\
r^{-4}&:&
\begin{cases}
\rho^o_3=-|\sigma^o|^2\\
\sigma^o_3=0
\end{cases}\\
r^{-5}&:&
\begin{cases}
\rho^o_4=0\\
\sigma^o_4=\sigma^o|\sigma^o|^2-\frac{1}{2}\Psi^o_0
\end{cases}\\
r^{-6}&:&
\begin{cases}
\rho^o_5=-|\sigma^o|^4+\frac{1}{3}\textrm{Re}(\bar{\sigma}^o\Psi^o_0)\\
\sigma^o_5=-\frac{1}{3}\Psi^1_0.
\end{cases}
\end{eqnarray}

(4) $D\Psi_1-\delta'\Psi_0=4\rho\Psi_1-4\alpha\Psi_0$.
\begin{eqnarray}
r^{-2}&:&(\Psi_1)^o_1=0\\
r^{-3}&:&(\Psi_1)^o_2=0\\
r^{-4}&:&(\Psi_1)^o_3=0\\
r^{-5}&:&0=0,\textrm{ i.e. }\Psi_1=\Psi^o_1r^{-4}+O(r^{-5})\\
r^{-6}&:&(\Psi_1)^o_5=-\eth'\Psi^o_0.
\end{eqnarray}

(5) $D\Psi_2-\delta'\Psi_1=3\rho\Psi_2-2\alpha\Psi_1+\sigma'\Psi_0$.
\begin{eqnarray}
r^{-2}&:& (\Psi_2)^o_1=0\\
r^{-3}&:&(\Psi_2)^o_2=0\\
r^{-4}&:&0=0,\textrm{ i.e. }\Psi_2=\Psi^o_2r^{-3}+O(r^{-4})\\
r^{-5}&:&(\Psi_2)^o_4=-\eth'\Psi^o_1-\sigma'^o_0\Psi^o_0.
\end{eqnarray}

(6) $D\Psi_3-\delta'\Psi_2=2\rho\Psi_3+2\sigma'\Psi_1$.
\begin{eqnarray}
r^{-2}&:& (\Psi_3)^o_1=0\\
r^{-3}&:&0=0,\textrm{ i.e. }\Psi_3=\Psi^o_3r^{-2}+O(r^{-3})\\
r^{-4}&:&(\Psi_3)^o_3=-\eth'\Psi^o_2-2\sigma'^o_0\Psi^o_1.
\end{eqnarray}

(7) $D\Psi_4-\delta'\Psi_3=\rho\Psi_4+2\alpha\Psi_3+3\sigma'\Psi_2$.
\begin{eqnarray}
r^{-2}&:&0=0,\textrm{ i.e. }\Psi_4=\Psi^o_4r^{-1}+O(r^{-2})\\
r^{-3}&:&(\Psi_4)^o_2=-\eth'\Psi^o_3-3\sigma'^o_0\Psi^o_2.
\end{eqnarray}
Ergo, asymptotically de Sitter spacetimes have the peeling property, arising from the fall-off of $\Psi_0$ being $O(r^{-5})$, such that
\begin{eqnarray}
\Psi_n=O(r^{n-5}),
\end{eqnarray}
where $n=0,1,2,3,4$.

(8) $\displaystyle D\rho'=\rho'\rho+\sigma'\sigma-\Psi_2-\frac{\Lambda}{3}$.
\begin{eqnarray}
1&:&\rho'^o_{-1}=-\frac{\Lambda}{6}\\
r^{-1}&:&0=0\\
r^{-2}&:&0=0\\
r^{-3}&:&\rho'^o_2=\Psi^o_2-\sigma^o\sigma'^o_1\\
r^{-4}&:&\rho'^o_3=-\frac{1}{2}(\rho^o_3\rho'^o_1+\rho^o_5\rho'^o_{-1}+\sigma'^o_0\sigma^o_4+\sigma'^o_2\sigma^o-(\Psi_2)^o_4).
\end{eqnarray}

(9) $D\omega=\rho\omega+\sigma\bar{\omega}+\alpha'-\bar{\alpha}$.
\begin{eqnarray}
r^{-1}&:&\alpha'^o_1=\bar{\alpha}^o_1\\
r^{-2}&:&\alpha'^o_2=\bar{\alpha}^o_2\\
r^{-3}&:&\omega^o_2=\bar{\alpha}^o_3-\alpha'^o_3-\sigma^o\bar{\omega}^o_1=-\frac{1}{2}\Psi^o_1-\sigma^o\bar{\omega}^o_1,\textrm{ from }(10)\textrm{ and }(11).
\end{eqnarray}

(10) $D\alpha=\alpha\rho-\alpha'\bar{\sigma}$.

(11) $D\alpha'=\alpha'\rho-\alpha\sigma-\Psi_1$.
\begin{eqnarray}
r^{-2}&:&
\begin{cases}
0=0\\
0=0
\end{cases}\\
r^{-3}&:&
\begin{cases}
\alpha^o_2=\bar{\alpha}^o\bar{\sigma}^o\\
\alpha'^o_2=\alpha^o\sigma^o
\end{cases}\\
r^{-4}&:&
\begin{cases}
\alpha^o_3=\alpha^o|\sigma^o|^2\\
\alpha'^o_3=\bar{\alpha}^o|\sigma^o|^2+\frac{1}{2}\Psi^o_1.
\end{cases}
\end{eqnarray}

(12) $\tau=\bar{\alpha}-\alpha'$.
\begin{eqnarray}
r^{-1}&:&\tau^o_1=0\\
r^{-2}&:&\tau^o_2=0\\
r^{-3}&:&\tau^o_3=-\frac{1}{2}\Psi^o_1.
\end{eqnarray}

(13) $\displaystyle D\gamma=\tau\alpha-\bar{\tau}\alpha'+\Psi_2-\frac{\Lambda}{6}$.
\begin{eqnarray}
1&:&\gamma^o_{-1}=-\frac{\Lambda}{6}\\
r^{-1}&:&0=0\\
r^{-2}&:&0=0\\
r^{-3}&:&\gamma^o_2=-\frac{1}{2}\Psi^o_2.
\end{eqnarray}

(14) $DU=2\textrm{Re}(\bar{\tau}\omega-\gamma)$.
\begin{eqnarray}
r&:&0=0\\
1&:&0=0\\
r^{-1}&:&0=0\\
r^{-2}&:&U^o_1=2\textrm{Re}(\gamma^o_2)=-\textrm{Re}(\Psi^o_2)\\
r^{-3}&:&U^o_2=\textrm{Re}(\gamma^o_3).
\end{eqnarray}

(15) $D\kappa'=\tau\sigma'+\bar{\tau}\rho'-\Psi_3$.
\begin{eqnarray}
r^{-2}&:&\kappa'^o_1=\Psi^o_3-\frac{\Lambda}{12}\bar{\Psi}^o_1\\
r^{-3}&:&2\kappa'^o_2=(\Psi_3)^o_3-\sigma'^o_0\tau^o_3-\bar{\tau}^o_4\rho'^o_{-1}.
\end{eqnarray}

(16) $D\xi^\theta=\rho\xi^3+\sigma\overline{\xi^\theta}$.

(17) $D\xi^\phi=\rho\xi^4+\sigma\overline{\xi^\phi}$.
\begin{eqnarray}
r^{-2}&:&
\begin{cases}
0=0\\
0=0
\end{cases}\\
r^{-3}&:&
\begin{cases}
(\xi^\theta)^o_2=-\sigma^o(\overline{\xi^\theta})^o\\
(\xi^\phi)^o_2=-\sigma^o(\overline{\xi^\phi})^o
\end{cases}\\
r^{-4}&:&
\begin{cases}
(\xi^\theta)^o_3=|\sigma^o|^2(\xi^\theta)^o\\
(\xi^\phi)^o_3=|\sigma^o|^2(\xi^\phi)^o
\end{cases}\\
r^{-5}&:&
\begin{cases}
(\xi^\theta)^o_4=\left(\frac{1}{6}\Psi^o_0-\sigma^o|\sigma^o|^2\right)(\overline{\xi^\theta})^o\\
(\xi^\phi)^o_4=\left(\frac{1}{6}\Psi^o_0-\sigma^o|\sigma^o|^2\right)(\overline{\xi^\phi})^o.
\end{cases}
\end{eqnarray}

\newpage

(18) $DX^\theta=2\textrm{Re}(\bar{\tau}\xi^\theta)$.

(19) $DX^\phi=2\textrm{Re}(\bar{\tau}\xi^\phi)$.
\begin{eqnarray}
r^{-2}&:&
\begin{cases}
(X^\theta)^o_1=0\\
(X^\phi)^o_1=0
\end{cases}\\
r^{-3}&:&
\begin{cases}
(X^\theta)^o_2=0\\
(X^\phi)^o_2=0
\end{cases}\\
r^{-4}&:&
\begin{cases}
(X^\theta)^o_3=\frac{1}{3}\textrm{Re}(\bar{\Psi}^o_1(\xi^\theta)^o)\\
(X^\phi)^o_3=\frac{1}{3}\textrm{Re}(\bar{\Psi}^o_1(\xi^\phi)^o).
\end{cases}
\end{eqnarray}

This concludes solving all 19 radial equations involving the $D=\partial/\partial r$ derivative. Next, we deal with the other 19 equations.

(20) $D'\omega-\delta U=(\rho'+2i\textrm{Im}(\gamma))\omega+\bar{\sigma}'\bar{\omega}-\bar{\kappa}'$.
\begin{eqnarray}
1&:&\kappa'^o_0=\frac{\Lambda}{3}\bar{\omega}^o_1\label{kappaprime}\\
r^{-1}&:&\kappa'^o_1=-\dot{\bar{\omega}}^o_1-\frac{\Lambda}{4}\bar{\Psi}^o_1+\eth'U^o_0-\frac{2\Lambda}{3}\bar{\sigma}^o\omega^o_1.
\end{eqnarray}
Incidentally, (15) and (20) imply that
\begin{eqnarray}
\Psi^o_3=-\dot{\bar{\omega}}^o_1-\frac{\Lambda}{6}\bar{\Psi}^o_1+\eth'U^o_0-\frac{2\Lambda}{3}\bar{\sigma}^o\omega^o_1.\label{Psithree}
\end{eqnarray}

(21) $\textrm{Im}(\delta'\omega)=\textrm{Im}(\rho'+(\alpha+\bar{\alpha}')\omega)$.
\begin{eqnarray}
r^{-1}&:&\textrm{Im}(\rho'^o_1)=0\\
r^{-2}&:&\textrm{Im}(\rho'^o_2)=\textrm{Im}(\eth'\omega^o_1).
\end{eqnarray}

\newpage

(22) $D'\sigma-\delta\tau=\rho\bar{\sigma}'+\sigma(\rho'+2\gamma+2i\textrm{Im}(\gamma))+2\alpha'\tau$.

(23) $\displaystyle D'\rho-\delta'\tau=\sigma\sigma'+\rho(\bar{\rho}'+2\textrm{Re}(\gamma))-2\alpha\tau-\Psi_2-\frac{\Lambda}{3}$.
\begin{eqnarray}
1&:&
\begin{cases}0=0\\
0=0
\end{cases}\\
r^{-1}&:&
\begin{cases}0=0\\
0=0
\end{cases}\\
r^{-2}&:&
\begin{cases}\sigma'^o_1=-\dot{\bar{\sigma}}^o\\
U^o_0=-\rho'^o_1-\frac{\Lambda}{6}|\sigma^o|^2
\end{cases}\\
r^{-3}&:&
\begin{cases}\sigma'^o_2=-\bar{\sigma}^o\rho'^o_1-\frac{\Lambda}{12}\bar{\Psi}^o_0\\
\phantom{\sigma'^o_2}=\bar{\sigma}^o\left(U^o_0+\frac{\Lambda}{6}|\sigma^o|^2\right)-\frac{\Lambda}{12}\bar{\Psi}^o_0\\
\rho'^o_2=\Psi^o_2+\sigma^o\dot{\bar{\sigma}}^o,\\
\textrm{in accordance to (8) as well.}
\end{cases}
\end{eqnarray}

(24) $D'\rho'-\delta\kappa'=\sigma'\bar{\sigma}'+\rho'(\rho'-2\textrm{Re}(\gamma))-2\alpha'\kappa'$.
\begin{eqnarray}
r^2&:&0=0\\
r&:&0=0\\
1&:&0=0\\
r^{-1}&:&\dot{\rho}'^o_1=\frac{\Lambda}{3}\frac{\partial}{\partial u}(|\sigma^o|^2)+\frac{\Lambda}{3}\textrm{Re}(\eth\bar{\omega}^o_1),\textrm{ giving}\\
&\ &\rho'^o_1=\Theta(\theta, \phi)+\frac{\Lambda}{3}|\sigma^o|^2+\frac{\Lambda}{3}\int{\textrm{Re}(\eth\bar{\omega}^o_1)du}.
\end{eqnarray}
Here, $\Theta(\theta, \phi)$ is an arbitrary real function. From (23), this gives
\begin{eqnarray}
\dot{U}^o_0&=&-\frac{\Lambda}{2}\frac{\partial}{\partial u}(|\sigma^o|^2)-\frac{\Lambda}{3}\textrm{Re}(\eth\bar{\omega}^o_1),\\
U^o_0&=&-\Theta(\theta, \phi)-\frac{\Lambda}{2}|\sigma^o|^2-\frac{\Lambda}{3}\int{\textrm{Re}(\eth\bar{\omega}^o_1)du}.
\end{eqnarray}

(25) $D'\sigma'-\delta'\kappa'=\sigma'(2\textrm{Re}(\rho')-2\gamma-2i\textrm{Im}(\gamma))+2\alpha\kappa'+\Psi_4$.
\begin{eqnarray}
r&:&0=0\\
1&:&0=0\\
r^{-1}&:&\Psi^o_4=-\ddot{\bar{\sigma}}^o-\frac{2\Lambda}{3}\bar{\sigma}^oU^o_0-\frac{\Lambda}{3}\eth'\bar{\omega}^o_1-\frac{\Lambda^2}{9}\bar{\sigma}^o|\sigma^o|^2+\frac{\Lambda^2}{36}\bar{\Psi}^o_0\\
&\ &\phantom{\Psi^o_4}=-\ddot{\bar{\sigma}}^o+\frac{2\Lambda}{3}\bar{\sigma}^o\Theta-\frac{\Lambda}{3}\eth'\bar{\omega}^o_1+\frac{2\Lambda^2}{9}\bar{\sigma}^o|\sigma^o|^2+\frac{2\Lambda^2}{9}\bar{\sigma}^o\int{\textrm{Re}(\eth\bar{\omega}^o_1)du}+\frac{\Lambda^2}{36}\bar{\Psi}^o_0.\ \ \ \ \ 
\end{eqnarray}

(26) $D'\xi^\theta-\delta X^\theta=(\rho'+2i\textrm{Im}(\gamma))\xi^\theta+\bar{\sigma}'\overline{\xi^\theta}$.

(27) $D'\xi^\phi-\delta X^\phi=(\rho'+2i\textrm{Im}(\gamma))\xi^\phi+\bar{\sigma}'\overline{\xi^\phi}$.
\begin{eqnarray}
1&:&
\begin{cases}0=0\\
0=0
\end{cases}\\
r^{-1}&:&
\begin{cases}(\dot{\xi}^\theta)^o=-\frac{\Lambda}{3}\sigma^o(\overline{\xi^\theta})^o\label{xitheta}\\
(\dot{\xi}^\phi)^o=-\frac{\Lambda}{3}\sigma^o(\overline{\xi^\phi})^o\label{xiphi}
\end{cases}\\
r^{-2}&:&
\begin{cases}0=0\\
0=0
\end{cases}\\
r^{-3}&:&
\begin{cases}0=0\\
0=0.
\end{cases}
\end{eqnarray}

(28) $D'\alpha-\delta'\gamma=\alpha(\bar{\rho}'-2i\textrm{Im}(\gamma))+(\bar{\alpha}-2\alpha')\sigma'-\rho\kappa'-\Psi_3$.

(29) $D'\alpha'+\delta\gamma=\alpha'(\rho'+2i\textrm{Im}(\gamma))-\alpha\bar{\sigma}'+\sigma\kappa'-\tau\rho'$.
\begin{eqnarray}
1&:&
\begin{cases}
0=0\\
0=0
\end{cases}\\
r^{-1}&:&
\begin{cases}
\dot{\alpha}^o=\frac{\Lambda}{3}\bar{\alpha}^o\bar{\sigma}^o+\frac{\Lambda}{6}\bar{\omega}^o_1\label{alpha}\\
0=0
\end{cases}\\
r^{-2}&:&
\begin{cases}0=0\\
0=0.
\end{cases}
\end{eqnarray}

(30) $\delta\rho-\delta'\sigma=(\bar{\alpha}-\alpha')\rho-(3\alpha+\bar{\alpha}')\sigma-\Psi_1$.
\begin{eqnarray}
r^{-3}&:&\omega^o_1=\eth'\sigma^o\\
r^{-4}&:&0=0.
\end{eqnarray}

(31) $\delta'\rho'-\delta\sigma'=(\bar{\alpha}'-\alpha)\rho'-(3\alpha'+\bar{\alpha})\sigma'-\Psi_3$.
\begin{eqnarray}
r^{-1}&:&0=0\\
r^{-2}&:&\Psi^o_3=-\eth\dot{\bar{\sigma}}^o-\frac{\Lambda}{6}\bar{\Psi}^o_1+\eth'U^o_0+\frac{\Lambda}{3}\sigma^o\eth'\bar{\sigma}^o.
\end{eqnarray}
Well,
\begin{eqnarray}
[\partial_u,\eth]\bar{\sigma}^o&=&\partial_u((\xi^\theta)^o\partial_\theta\bar{\sigma}^o+(\xi^\phi)^o\partial_\phi\bar{\sigma}^o-4\bar{\alpha}^o\bar{\sigma}^o)-\eth\dot{\bar{\sigma}}^o\label{com1}\\
&=&(\dot{\xi}^\theta)^o\partial_\theta\bar{\sigma}^o+(\dot{\xi}^\phi)^o\partial_\phi\bar{\sigma}^o-4\dot{\bar{\alpha}}^o\bar{\sigma}^o+\eth\dot{\bar{\sigma}}^o-\eth\dot{\bar{\sigma}}^o\\
&=&-\frac{\Lambda}{3}\sigma^o((\overline{\xi^\theta})^o\partial_\theta\bar{\sigma}^o+(\overline{\xi^\phi})^o\partial_\phi\bar{\sigma}^o+4\alpha^o\bar{\sigma}^o)-\frac{2\Lambda}{3}\bar{\sigma}^o\eth'\sigma^o\\
&=&-\frac{\Lambda}{3}\sigma^o\eth'\bar{\sigma}^o-\frac{2\Lambda}{3}\bar{\sigma}^o\eth'\sigma^o.\label{com2}
\end{eqnarray}
So,
\begin{eqnarray}
\Psi^o_3=-\partial_u(\eth\bar{\sigma}^o)-\frac{\Lambda}{6}\bar{\Psi}^o_1+\eth'U^o_0-\frac{2\Lambda}{3}\bar{\sigma}^o\eth'\sigma^o,
\end{eqnarray}
as expected from (15) and (20) in Eq. (\ref{Psithree}).

(32) $\textrm{Im}(\delta'\xi^\theta)=\textrm{Im}((\alpha+\bar{\alpha}')\xi^\theta)$.

(33) $\textrm{Im}(\delta'\xi^\phi)=\textrm{Im}((\alpha+\bar{\alpha}')\xi^\phi)$.
\begin{eqnarray}
r^{-2}&:&
\begin{cases}\textrm{Im}(\delta'^o(\xi^\theta)^o)=2\textrm{Im}(\alpha^o(\xi^\theta)^o)\label{1}\\
\textrm{Im}(\delta'^o(\xi^\phi)^o)=2\textrm{Im}(\alpha^o(\xi^\phi)^o)\label{2}.
\end{cases}
\end{eqnarray}

(34) $\displaystyle \delta\alpha+\delta'\alpha'=\alpha\bar{\alpha}+\alpha'\bar{\alpha}'+2\alpha\alpha'-\rho\rho'+\sigma\sigma'-\Psi_2+\frac{\Lambda}{6}$.
\begin{eqnarray}
1&:&0=0\\
r^{-1}&:&0=0\\
r^{-2}&:&2\textrm{Re}(\delta^o\alpha^o)-4|\alpha^o|^2=\Theta+\frac{\Lambda}{3}\int{\textrm{Re}(\eth^2\bar{\sigma}^o)du}.\label{Gauss}
\end{eqnarray}

(35) $D'\Psi_0-\delta\Psi_1=(4\gamma+\rho')\Psi_0-2(2\tau-\alpha')\Psi_1+3\sigma\Psi_2$.
\begin{eqnarray}
r^{-4}&:&0=0\\
r^{-5}&:&\dot{\Psi}^o_0=\eth\Psi^o_1+3\sigma^o\Psi^o_2+\frac{\Lambda}{6}\Psi^1_0.
\end{eqnarray}

(36) $D'\Psi_1-\delta\Psi_2=-\kappa'\Psi_0+2(\gamma+\rho')\Psi_1-3\tau\Psi_2+2\sigma\Psi_3$.
\begin{eqnarray}
r^{-3}&:&0=0\\
r^{-4}&:&\dot{\Psi}^o_1=\eth\Psi^o_2+2\sigma^o\Psi^o_3-\frac{\Lambda}{6}\eth'\Psi^o_0.
\end{eqnarray}

(37) $D'\Psi_2-\delta\Psi_3=-2\kappa'\Psi_1+3\rho'\Psi_2-2(\tau+\alpha')\Psi_3+\sigma\Psi_4$.
\begin{eqnarray}
r^{-2}&:&0=0\\
r^{-3}&:&-\partial_u(\Psi^o_2+\sigma^o\dot{\bar{\sigma}}^o)=-|\dot{\sigma}^o|^2-\eth\Psi^o_3-\frac{2\Lambda}{3}\Theta|\sigma^o|^2+\frac{\Lambda}{3}\sigma^o\eth'\eth\bar{\sigma}^o+\frac{\Lambda}{6}\eth'\Psi^o_1\nonumber\\&\ &\phantom{-\partial_u(\dot{\Psi}^o_2+\sigma^o\bar{\sigma}^o)=}-\frac{2\Lambda^2}{9}|\sigma^o|^4-\frac{2\Lambda^2}{9}|\sigma^o|^2\int{\textrm{Re}(\eth^2\bar{\sigma}^o)du}-\frac{\Lambda^2}{18}\textrm{Re}(\bar{\sigma}^o\Psi^o_0).\
\end{eqnarray}

(38) $D'\Psi_3-\delta\Psi_4=-3\kappa'\Psi_2-2(\gamma-2\rho')\Psi_3-(\tau+4\alpha')\Psi_4$.
\begin{eqnarray}
r^{-1}&:&0=0\\
r^{-2}&:&\dot{\Psi}^o_3=\eth\Psi^o_4-\frac{\Lambda}{6}\eth'\Psi^o_2+\frac{\Lambda^2}{18}\bar{\sigma}^o\Psi^o_1,\textrm{ giving }0=0.
\end{eqnarray}
In showing that the above is identically satisfied, we need the following four commutators:
\begin{eqnarray}
[\partial_u,\eth']U^o_0&=&-\frac{\Lambda}{3}\bar{\sigma}^o\eth U^o_0\\
{[}\partial_u,\eth]\dot{\bar{\sigma}}^o&=&-\frac{\Lambda}{3}\sigma^o\eth'\dot{\bar{\sigma}}^o-\frac{2\Lambda}{3}\dot{\bar{\sigma}}^o\eth'\sigma^o\\
{[}\partial_u,\eth']\bar{\sigma}^o&=&\frac{\Lambda}{3}\bar{\sigma}^o\eth\bar{\sigma}^o\\
{[}\eth,\eth']\eth\bar{\sigma}^o&=&2(2\textrm{Re}(\delta^o\alpha^o)-4|\alpha^o|^2)\eth\bar{\sigma}^o\\
&=&2\left(\Theta+\frac{\Lambda}{3}\int{\textrm{Re}(\eth^2\bar{\sigma}^o)du}\right)\eth\bar{\sigma}^o.
\end{eqnarray}
The derivation of the first three of these commutators is similar to that for $[\partial_u,\eth]\bar{\sigma}^o$, as was done in (31) (Eqs. (\ref{com1})-(\ref{com2})). The derivation of the $[\eth,\eth']\eth\bar{\sigma}^o$ commutator involves (32) and (33), viz. Eq. (\ref{1}) \footnote{This is consistent with (4.14.1) in Penrose-Rindler \cite{Pen87} for $[\eth,\eth']\eth\bar{\sigma}^o$, where $\eth\bar{\sigma}^o$ is a quantity with $p=-2$ and $q=0$. The term involving $\textrm{Im}(\rho')$\th\ is zero because $\gamma'=0$ gives \th$\ =\partial/\partial r$, but $\eth\bar{\sigma}^o$ is independent of $r$. The term involving $\textrm{Im}(\rho)$\th$'$ is zero because $\rho$ is real.}. Note that all these commutators (including the one used in (31)) are zero when $\Lambda=0$, except $[\eth,\eth']\eth\bar{\sigma}^o=2\Theta\eth\bar{\sigma}^o$.

In the case of de Sitter spacetime where $\sigma^o=0$, we have $U^o_0=-1/2$. Hence we take $\Theta=1/2$.

\begin{acknowledgments}
I wish to thank J\"{o}rg Frauendiener for very helpful discussions which clarified several subtle technical details, thereby enhancing the manuscript. The suggestions by the reviewer towards improving the manuscript are also very much appreciated.
\end{acknowledgments}

\bibliographystyle{spphys}       
\bibliography{Citation}

\end{document}